\documentstyle[12pt,aaspp4]{article}

\begin{document}

\title{PROPERTIES OF THE TRANS-NEPTUNIAN BELT: STATISTICS FROM THE
CFHT SURVEY\footnotemark}
\author{Chadwick A. Trujillo\altaffilmark{2}}
\affil{Institute for Astronomy, 2680 Woodlawn Drive, Honolulu, HI
96822 \\ chad@ifa.hawaii.edu}
\author{David C. Jewitt}
\affil{Institute for Astronomy, 2680 Woodlawn Drive, Honolulu, HI
96822 \\ jewitt@ifa.hawaii.edu}
\and
\author{Jane X. Luu}
\affil{Leiden Observatory, PO Box 9513, 2300 RA Leiden, The
Netherlands \\ luu@strw.leidenuniv.nl}
\footnotetext{Based on observations collected at
Canada-France-Hawaii Telescope, which is operated by the National
Research Council of Canada, the Centre National de la Recherche Scientifique
de France, and the University of Hawaii.}
\altaffiltext{2}{Now at California Institute of Technology, MS 150-21,
Pasadena, CA 91125. chad@gps.caltech.edu}

\begin{abstract}
  We present the results of a wide-field survey designed to measure
  the size, inclination, and radial distributions of Kuiper Belt
  Objects (KBOs).  The survey found 86 KBOs in 73 square degrees
  observed to limiting red magnitude 23.7 using the
  Canada-France-Hawaii Telescope and the 12k x 8k CCD Mosaic camera.
  For the first time, both ecliptic and off-ecliptic fields were
  examined to more accurately constrain the inclination distribution
  of the KBOs.  The survey data were processed using an automatic
  moving object detection algorithm, allowing a careful
  characterization of the biases involved.  In this work, we quantify
  fundamental parameters of the Classical KBOs (CKBOs), the most
  numerous objects found in our sample, using the new data and a
  maximum likelihood simulation.  Deriving results from our best-fit
  model, we find that the size distribution follows a differential
  power law with exponent $q = 4.0^{+0.6}_{-0.5}$ ($1 \sigma$, or
  68.27\% confidence).  In addition, the CKBOs inhabit a very thick
  disk consistent with a Gaussian distribution of inclinations with a
  Half-Width of $i_{1/2} = 20\arcdeg^{+6\arcdeg}_{-4\arcdeg}$ (1
  $\sigma$).  We estimate that there are $N_{\rm CKBOs}(D > 100 \mbox
  { km}) = 3.8^{+2.0}_{-1.5} \times 10^{4}$ ($1\sigma$) CKBOs larger
  than 100 km in diameter.  We also find compelling evidence for an
  outer edge to the CKBOs at heliocentric distance $R = 50$ AU.
\end{abstract}


\section{Introduction}
\label{intro}
The rate of discovery of Kuiper Belt Objects (KBOs) has increased
dramatically since the first member (1992 $\rm QB_{1}$) was found
(Jewitt \& Luu 1993).  As of Dec 2000, $\sim 400$ KBOs are known.
These bodies exist in three dynamical classes (Jewitt, Luu \& Trujillo
1998): (1) the Classical KBOs (CKBOs) occupy nearly circular
(eccentricities $e < 0.25$) orbits with semimajor axes $41 \mbox{ AU}
\lesssim a \lesssim 46 \mbox{ AU}$, and they constitute $\sim 70$ \%
of the observed population; (2) the Resonant KBOs occupy mean-motion
resonances with Neptune, such as the 3:2 (the Plutinos, $a \approx
39.4$ AU) and 2:1 ($a \approx 47.8$ AU), and comprise $\sim 20$ \% of
the known objects; (3) the Scattered KBOs represent only $\sim 10$ \%
of the known KBOs, but possess the most extreme orbits, with median
semimajor axis $a \sim 90$ AU and eccentricity $e \sim 0.6$,
presumably due to a weak interaction with Neptune (Duncan \& Levison
1997, Luu et al.  1997, and Trujillo, Jewitt \& Luu 2000).  Although
these classes are now well established, only rudimentary information
has been collected about their populations.  One reason is that only a
fraction of the known KBOs were discovered in well-parametrized
surveys that have been published in the open literature (principally
Jewitt \& Luu 1993 (1 KBO); Jewitt \& Luu 1995 (17 KBOs); Irwin,
Tremaine \& \.{Z}ytkow 1995 (2 KBOs); Jewitt, Luu \& Chen 1996 (15
KBOs); Gladman et al.  1998 (5 KBOs); Jewitt, Luu \& Trujillo 1998 (13
KBOs); and Chiang \& Brown 1999 (2 KBOs)).  In this work, we
characterize the fundamental parameters of the CKBOs: the size
distribution, inclination distribution and radial distribution using a
large sample (86 KBOs) discovered in a well-characterized survey.

The quintessential measurement of the size distribution relies on the
Cumulative Luminosity Function (CLF).  The CLF describes the number of
KBOs $\mbox{deg}^{-2}$ ($\Sigma$) near the ecliptic as a function of
apparent red magnitude ($m_{R}$).  It is fitted with the relation
$\log(\Sigma) = \alpha (m_{R} - m_{0})$, where $m_{0}$ is the red
magnitude at which $\Sigma = 1$ KBO $\mbox{deg}^{-2}$.  The slope
($\alpha$) is related to the size distribution (described later).
Although many different works have considered the CLF, two papers are
responsible for discovering the majority of KBOs found in published
surveys: Jewitt, Luu \& Chen (1996) and Jewitt, Luu \& Trujillo
(1998).  The former constrained the CLF over a 1.6 magnitude range
($23.2 < m_{R} < 24.8$) with 15 discovered KBOs while the latter
covered a complimentary 2.5 magnitude range ($20.5 < m_{R} < 23.0$),
discovering 13 objects.  Jewitt, Luu \& Trujillo (1998) measured the
CLF produced from these two data sets and found $\alpha = 0.58 \pm
0.05$ and $m_{0} = 23.27 \pm 0.11$.  Gladman et al. (1998) criticized
this work on 2 main counts: (1) they believed that Jewitt, Luu \& Chen
(1996) underestimated the number of KBOs and (2) the fit in the
Jewitt, Luu \& Trujillo (1998) survey used a least-squares approach
that assumed Gaussian errors rather than Poissonian errors.  Gladman
et al. (1998) found 5 additional KBOs and re-analysed the CLF using a
Poissonian maximum likelihood method to refit the CLF to (1) the
Jewitt, Luu \& Trujillo (1998) data without the Jewitt, Luu \& Chen
(1996) data and (2) a fit to the 6 different surveys available at the
time except for Tombaugh (1961), Kowal (1989) and Jewitt, Luu \& Chen
(1996).  Both of these fits were steeper but formally consistent with
the original Jewitt, Luu \& Trujillo (1998) data at the $\sim 1.5
\sigma$ level: (1) $\alpha = 0.72^{+0.30}_{-0.26}$ and $m_{0} =
23.3^{+0.2}_{-0.4}$ and (2) $\alpha = 0.76^{+0.10}_{-0.11}$ and $m_{0}
= 23.40^{+0.20}_{-0.18}$.  Chiang \& Brown (1999) find a flatter size
distribution of $\alpha = 0.52 \pm 0.05$ and $m_{0} = 23.5 \pm 0.06$
much closer to the Jewitt, Luu \& Trujillo (1998) result.  They
observed that the steep size distribution reported by Gladman et
al. (1998) was an artifact of their selective exclusion of part of the
available survey data, not of their use of a different fitting method.
The first goal of this work is to measure the CLF and additionally
constrain the power-law slope of the size distribution using a single
well-characterized survey and a maximum likelihood simulation which
allows for the correction of observational biases.

An accurate characterization of the inclination distribution of the
KBOs is critical to understanding the dynamical history of the outer
Solar System since the era of planetesimal formation.  We expect that
the KBOs formed by accretion in a very thin disk of particles with a
small internal velocity dispersion (e.g. Kenyon \& Luu 1998 and Hahn
\& Malhotra 1999) and a correspondingly small inclination
distribution.  However, the velocity dispersion indicated by the
inclination distribution in the present-day Kuiper Belt is large.
Jewitt, Luu \& Chen (1996) measured the {\it apparent} Half-Width of
the Kuiper Belt inclination distribution to be $\sim 5^{\circ}$.  They
noted a strong bias against observing high inclination objects in
ecliptic surveys, and they estimated the true distribution to be much
thicker, with an inclination distribution Half-Width of $\gtrsim
15^{\circ}$, corresponding to a vertical velocity dispersion of $\sim
1$ km/s.  Several conjectures have been advanced to explain the
thickness of the Kuiper Belt: Earth-mass planetesimals may have been
scattered through the belt in the late stages of the planet-formation
era, exciting the Kuiper Belt (Morbidelli \& Valsecchi 1997 and Petit
et al. 1999); stellar encounters may have enhanced the velocity
dispersion of the distant KBOs (Ida, Larwood \& Burkert, 2000); and
the dispersion velocity of small bodies tends to grow to roughly equal
the escape speed of the bodies contributing the most mass (the large
bodies for size distributions with $q < 4$) in the belt (Aarseth, Lin
\& Palmer 1993).  As there is much speculation about the origin of the
large velocity dispersion of the Kuiper Belt, but only one published
measurement (Jewitt, Luu \& Chen 1996), the second goal of this work
is to accurately quantify the inclination distribution from our large
sample of objects.

The radial extent of the Classical Kuiper Belt has not been well
constrained.  None of the CKBOs have been discovered beyond $R \approx
50$ AU.  This trend was first noted by Dones (1997) who suggested that
the 50--75 AU region may be depleted; he found the results of a
Monte-Carlo simulation of CKBOs drawn from a rather flat differential
size distribution (power-law index $q = 3$) to be inconsistent with
the observations of the 6 CKBOs discovered by Jewitt, Luu \& Chen
(1996).  Jewitt, Luu \& Trujillo (1998) discovered all of their KBOs
at heliocentric distances $R < 46$ AU.  In the absence of other
effects, one should expect to find fewer bodies with $R > 50$ AU than
with $R \sim 40$ AU, as the former are about a magnitude fainter than
the latter.  However, through the use of a Monte-Carlo model they
demonstrated that the bias against objects beyond 50 AU is not strong
enough to explain the distribution of discovery distances.  They
speculated that the lack of bodies discovered beyond 50 AU could be
caused by a combination of (1) a decrease in the maximum KBO size (and
reduction in the brightest and most detectable objects) beyond 50 AU
or (2) the size distribution might steepen beyond 50 AU, putting more
of the mass in the smaller, less-detectable bodies.  They also
suggested that the lack of $R>50$ AU objects could be explained by an
outer edge to the Classical Kuiper Belt at 50 AU.

Two later papers questioned the existence of an edge to the Kuiper
Belt near 50 AU.  Gladman et al. (1998) suggested that the number of
objects expected to be discovered beyond 50 AU is highly dependent on
the size distribution because steep size distributions reduce the
number of large (bright) bodies relative to small (faint) bodies.
Gladman et al. (1998) adopted a relatively steep distribution
($q=4.65$), and found no significant evidence of a truncated belt.
Chiang \& Brown (1999) found that 8\%--13\% of the $\sim 100$ objects
known at the time should have been found beyond 50 AU, and suggested
that this precludes the presence of a density enhancement beyond 50
AU, but could not definitively rule out a density deficit.  Allen,
Bernstein \& Malhotra (2001) have also recently reported the detection
of an outer edge to the Kuiper Belt.  The third goal of the present
work is to test the distribution of the discovery distances for the
presence of an outer edge to the Kuiper Belt.

\section{Survey Data}

Observations were made at the 3.6 m diameter Canada-France-Hawaii
Telescope using the 12288 x 8192, 15 {\micron} pixel mosaic CCD (CFHT
12k; Cuillandre et al. 2000).  Built at the University of Hawaii (UH)
the CFHT 12k comprises 12 edge-abutted, thinned, high quantum
efficiency (QE $\sim 0.75$), 4096 x 2048 pixel Lincoln Laboratory
CCDs.  It is currently the largest close-packed CCD camera in the
world.  When mounted at the CFHT f/4 prime focus the camera yields a
plate scale of 0.206 arc sec/pixel, corresponding to a 0.330 sq deg
field of view in each 200 Mbyte image.  Images were taken through a
Mould $R$ filter, with a central wavelength of 6581 {\AA} and a
bandwidth of 1251 {\AA}.  Instrumental parameters of the survey are
summarized in Table~\ref{obs}.

\begin{figure}
\plotfiddle{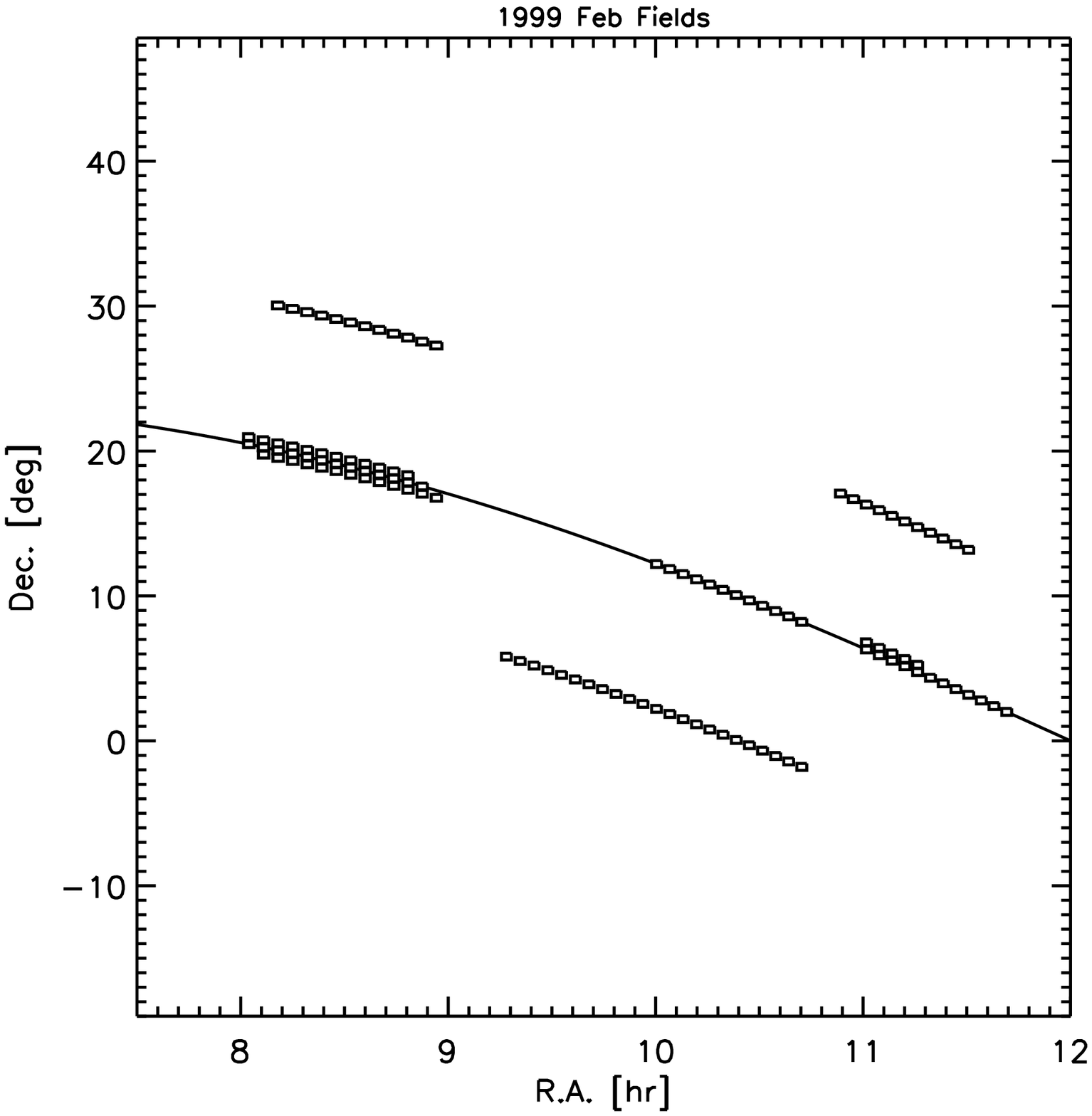}{7in}{0}{80}{80}{-250}{-50}
\figcaption{Fields imaged in Feb 1999.  The
ecliptic is denoted by a solid line.\label{feb99cfht}}
\end{figure}

\begin{figure}
\plotfiddle{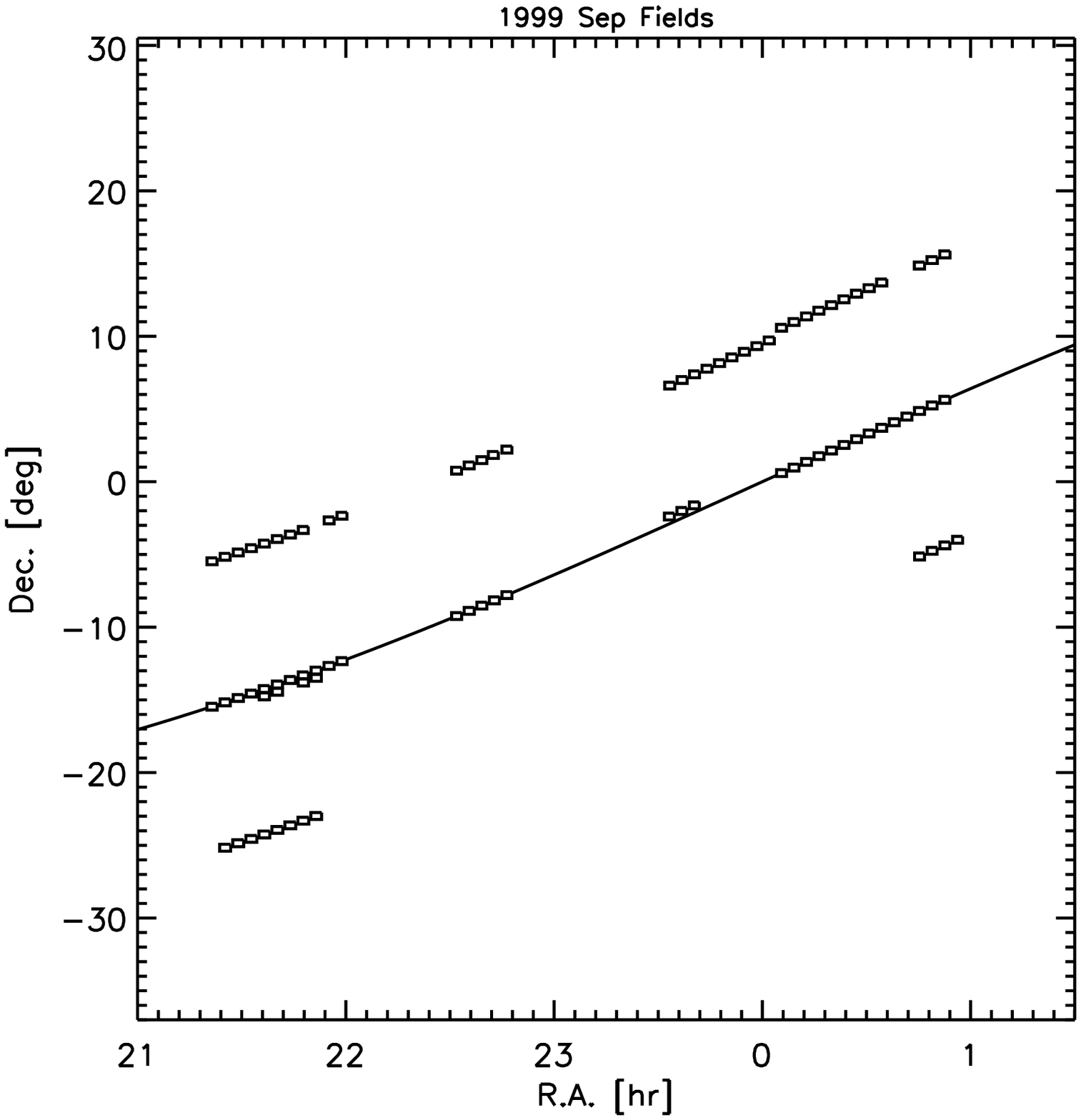}{7in}{0}{80}{80}{-250}{-50}
\figcaption{Fields imaged in Sep 1999. The ecliptic is denoted by a solid
line. \label{sep99cfht}}
\end{figure}

\begin{figure}
\plotfiddle{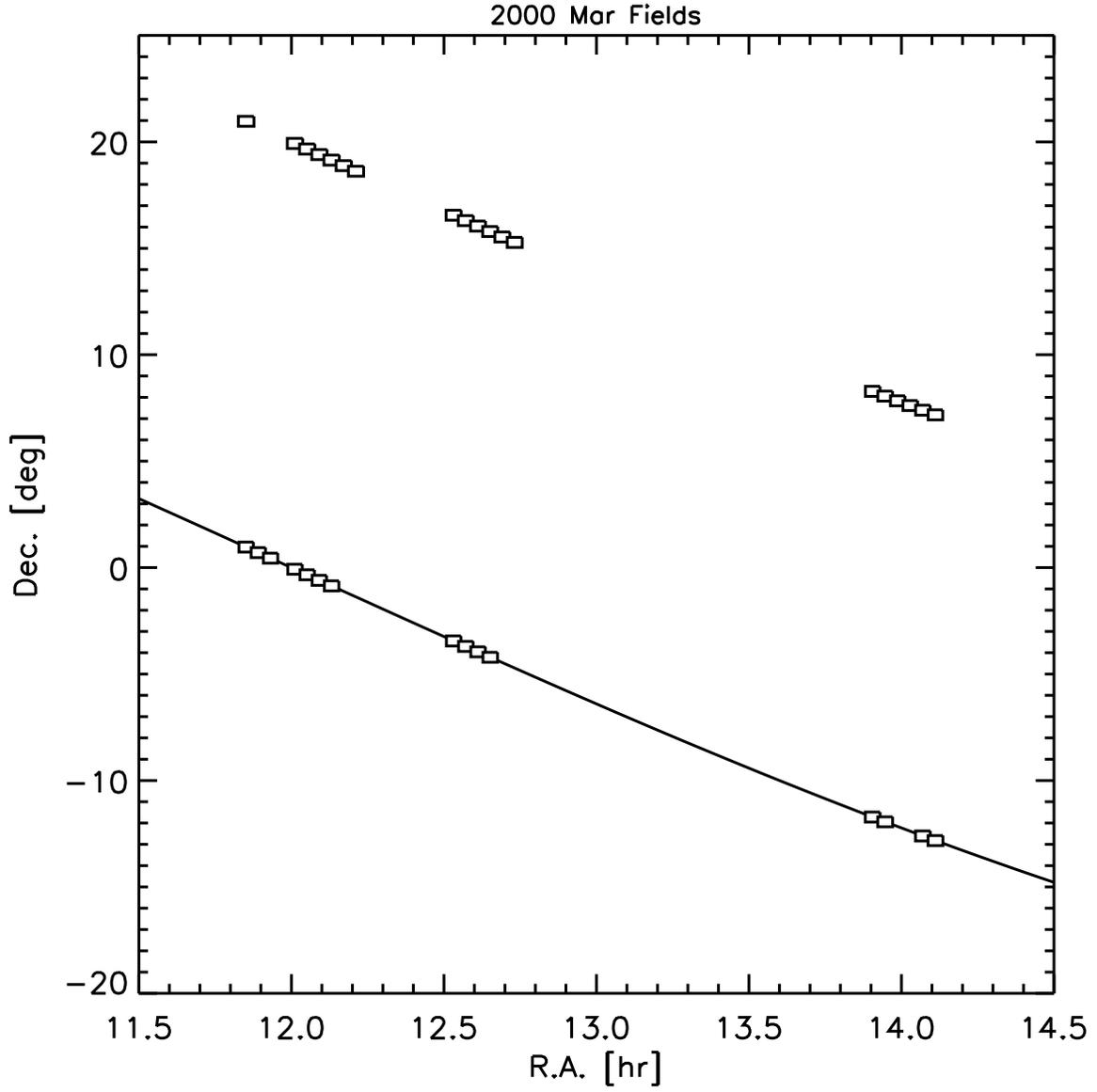}{7in}{0}{80}{80}{-250}{-50}
\figcaption{Fields imaged in Mar 2000. The ecliptic is denoted by a
solid line. \label{mar00cfht}}
\end{figure}

Observations were taken within a few days of new moon under
photometric conditions during three periods: Feb 10 -- 15 1999, Sep
5--8 1999, and Mar 31 -- Apr 3 2000.  Fields were imaged at airmasses
$< 1.7$ and were within 1.5 hours of opposition.  We chose to use
short 180 s exposures at the CFHT to maximize area coverage and
detection statistics.  All discovered objects were accessible for
recovery at the UH 2.2 m telescope during comparable seeing conditions
with exposure times of $< 600$ seconds.  Each field was imaged three
times (a ``field triplet''), with about 1 hour timebase between
exposures.  Fields imaged appear in Figures \ref{feb99cfht},
\ref{sep99cfht} and \ref{mar00cfht}, and in
Table~\ref{cfhtfieldtable}.  The CFHT observations were taken at three
ecliptic latitudes $\beta = 0^{\circ}$, $10^{\circ}$, and $20^{\circ}$
to probe the inclination distribution of the KBOs (see \S\ref{idist}).

Photometric calibrations were obtained from Landolt (1992) standard
stars imaged several times on each chip.  Three CFHT 12k chips of poor
quality were replaced between the Feb 1999 and Sep 1999 runs.  The
positions of four other CFHT 12k chips within the focal plane array
were changed to move the cosmetically superior chips towards the
center of the camera.  The photometric calibration accounts for these
changes, as shown in Table~\ref{cfhtphot}, containing the measured
photometric zero points of the chips.  In addition, chip 6 was not
used in Feb 1999 because of its extremely poor cosmetic quality.  The
area covered in the fields from Feb 1999 was corrected for this 8\%
reduction in field-of-view.  The area imaged in Mar 2000 included some
small field overlap (6\%), resulting in a minor correction applied to
the reported total area surveyed.

Each of the 12 CCDs in the CFHT 12k functions as an individual
detector, with its own characteristic bias level, flat field, gain
level, and orientation (at the $\sim 1^{\circ}$ level).  The bias
level for each chip was estimated using the row-by-row median of the
overscan region.  Flat fields were constructed from a combination of
(1) the median of normalized bias-subtracted twilight flat fields and
(2) a median of bias-subtracted data frames, with a clipping algorithm
used to remove excess counts due to bright stars.  Fields were
analysed by subtracting the overscan region, dividing by the composite
flats and searching for moving objects using our Moving Object
Detection Software (MODS, Trujillo \& Jewitt 1998).  We rejected bad
pixels through the use of a bad pixel mask.

Artificial moving objects were added to the data to quantify the
sensitivity of the moving object detection procedure (Trujillo \&
Jewitt 1998).  The seeing during the survey typically varied from 0.7
arc sec to 1.1 arc sec (FWHM).  Accordingly, we subdivided and
analysed the data in 3 groups based on the seeing.  Artificial moving
objects were added to bias-subtracted twilight sky-flattened images,
with profiles matched to the characteristic point-spread function for
each image group.  These images were then passed through the data
analysis pipeline.  The detection efficiency was found to be uniform
with respect to sky-plane speed in the 1 -- 10 arc sec/hr range. At
opposition, the apparent speed in arc sec/hr, $\dot{\theta}$, of an
object is dominated by the parallactic motion of the Earth, and
follows
\begin{equation}
\label{thetaeq}
\dot{\theta} \approx 148 \left(\frac{1-R^{-0.5}}{R-1}\right),
\end{equation}
where $R$ is heliocentric distance in AU (Luu \& Jewitt 1988).  From
Equation~\ref{thetaeq}, our speed limit criterion for the survey, $1
\mbox{ \arcsec} \mbox{ hr}^{-1} < \dot{\theta} < 10 \mbox{ \arcsec}
\mbox{ hr}^{-1}$, corresponds to opposition heliocentric distances $10
\mbox{ AU} \lesssim R \lesssim 140 \mbox{ AU}$, with efficiency
variations within this range due only to object brightness and seeing.

The magnitude-dependent efficiency function was fitted by
\begin{equation}
\label{cfhtefffunc}
\varepsilon =  \frac{\varepsilon_{\rm max}}{2} \left( \tanh \left( \frac{m_{R50} -
m_{R}}{\sigma} \right) + 1 \right) ,
\end{equation}
where $0 < \varepsilon < 1$ is the efficiency with which objects of
red magnitude $m_{R}$ are detected, $\varepsilon_{\rm max}$ is the
maximum efficiency, $m_{R50}$ is the magnitude at which $\varepsilon =
\varepsilon_{\rm max}/2$, and $\sigma$ magnitudes is the
characteristic range over which the efficiency drops from
$\varepsilon_{\rm max}$ to zero.  Table~\ref{effs} shows the
efficiency function derived for each seeing category, along with an
average of the seeing cases, weighted by sky area imaged, applicable
to the entire data set.  The efficiency function is known to greater
precision than the $\sim 0.1$ magnitude uncertainty on our discovery
photometry.  Changes to the efficiency function of $< 0.1$ magnitudes
produce no significant variation in our results for the size or
inclination distributions.

The MODS software, running on two Ultra 10 computers, was fast enough
to efficiently search for the KBOs in near real-time, so that newly
detected objects could be quickly discovered and re-imaged.  We imaged
$\sim 35$ field triplets each night at the CFHT, corresponding to
$\sim 20$ Gbytes of raw data collected per night, plus several more
Gbytes for flat fields and standard stars.  Eighty-six KBOs were found
in the CFHT survey, 2 of which were serendipitous re-detections of
known objects.  The discovery conditions of the detected objects
appear in Table \ref{cfhtdisc}.  Photometry was performed using a 2.5
arc second diameter synthetic aperture for discovery data, resulting
in median photometric error of 0.15 magnitudes and a maximum
photometric error of 0.3 for the faintest objects.  Our results are
unaffected by this error; randomly introducing $\pm 0.15$ magnitude
errors in our simulations (described later), and $\pm 0.3$ magnitude
errors in the faintest objects produced no statistically significant
change.  Trailing loss was insignificant as the KBOs moved only 0.15
arc sec during our integration.

\subsection{Recovery Observations and Orbits}
\label{recovery}

Extensive efforts were made to recover all objects using the UH 2.2 m
telescope.  Attempts were made to recover the objects one week after
discovery, then one, two and three months after discovery.  Most of
these attempts were successful, as demonstrated by the fact that 79 of
the 86 CFHT objects were recovered.  The loss of 7 objects is the
result of unusually poor weather during the Mar--May 1999 recovery
period.  Only 6 of the 79 recovered objects have arc-lengths shorter
than 30 days as of Dec 1, 2000.  Second opposition observations have
been acquired for 36 of the 78 KBOs found in Feb 1999 and Sep 1999.

Orbits derived from the discovery and recovery data appear in
Table~\ref{cfhtorbits}.  The listed elements are those computed by
Brian Marsden of the Minor Planet Center.  We also benefited from
orbital element calculations by David Tholen (Univ. of Hawaii).  Both
sources produced comparable orbital solutions to the astrometric data.

\begin{figure}
\plotfiddle{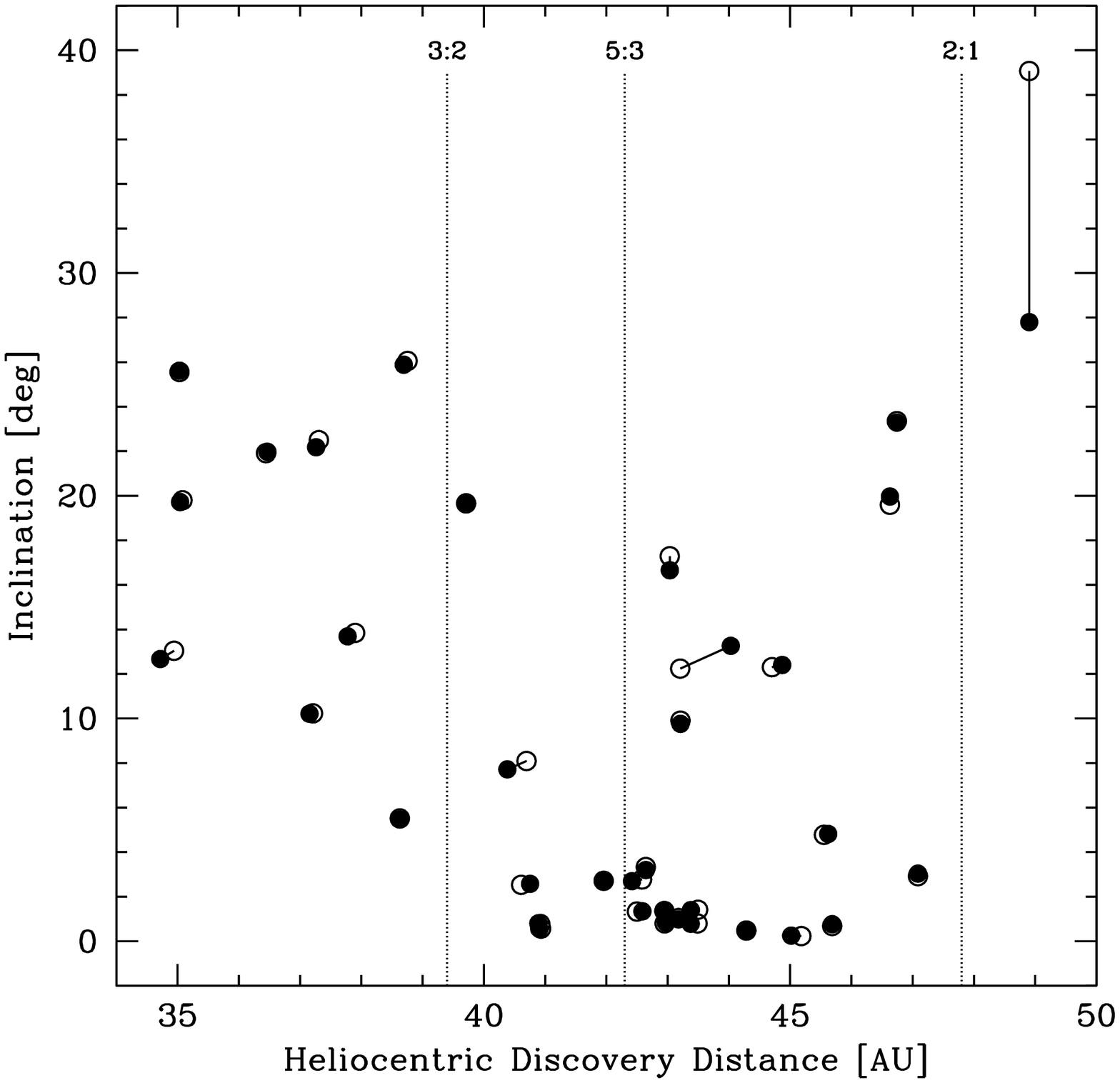}{6in}{0}{65}{65}{-200}{-50}
\figcaption{Inclination vs. discovery distance of all multi-opposition
KBOs.  The hollow circles represent quantities determined from $< 90$
day timebase during the first opposition.  The connected filled
circles represent the orbital solution including second opposition
observations.  Note that for all objects except one, quantities are
well determined during the first opposition. \label{ivsr-opp1}}
\end{figure}

\begin{figure}
\plotfiddle{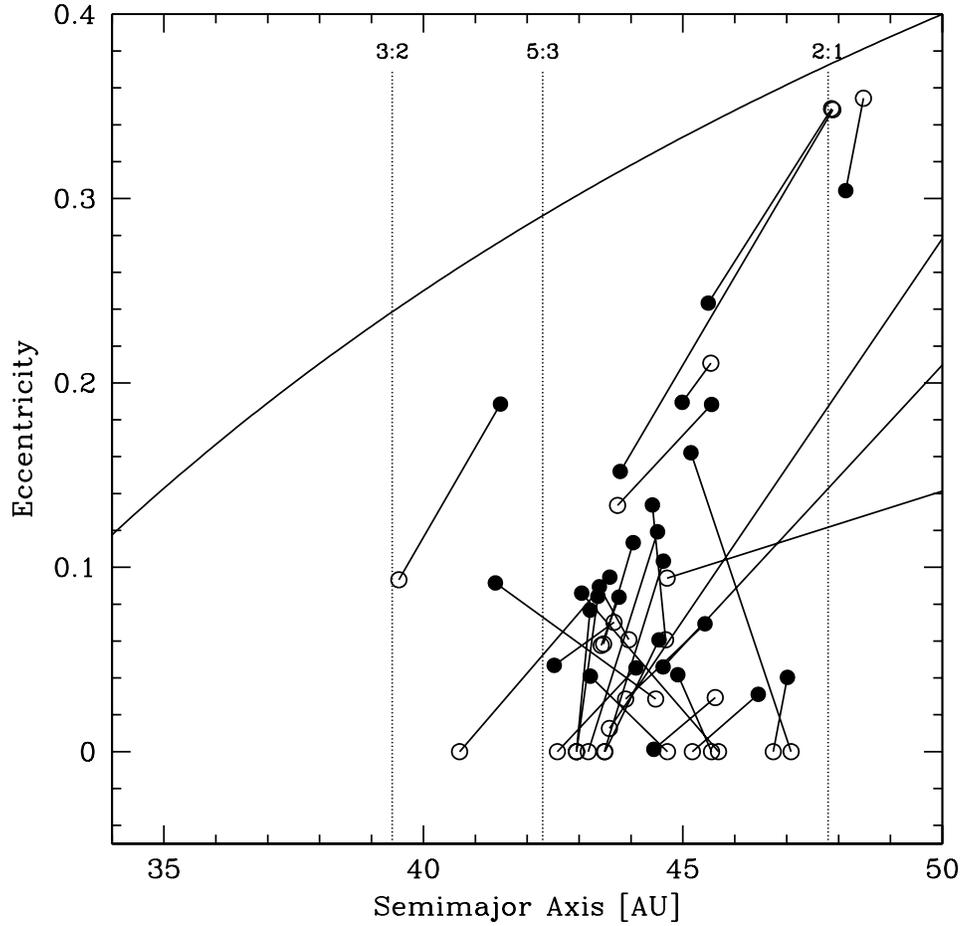}{6in}{0}{65}{65}{-200}{-50}
\figcaption{Eccentricity vs. semimajor axis of all multi-opposition
KBOs with $a < 50$ AU.  The hollow circles represent the orbits
determined during the first opposition.  The connected filled circles
show the orbital elements computed including second opposition
observations.  2 CKBOs were reclassified as Scattered KBOs, and 1
Scattered KBO was reclassified as a CKBO.  In addition, 3 Resonant
KBOs were reclassified as non-resonant objects. \label{evsa-opp1}}
\end{figure}

\begin{figure}
\plotfiddle{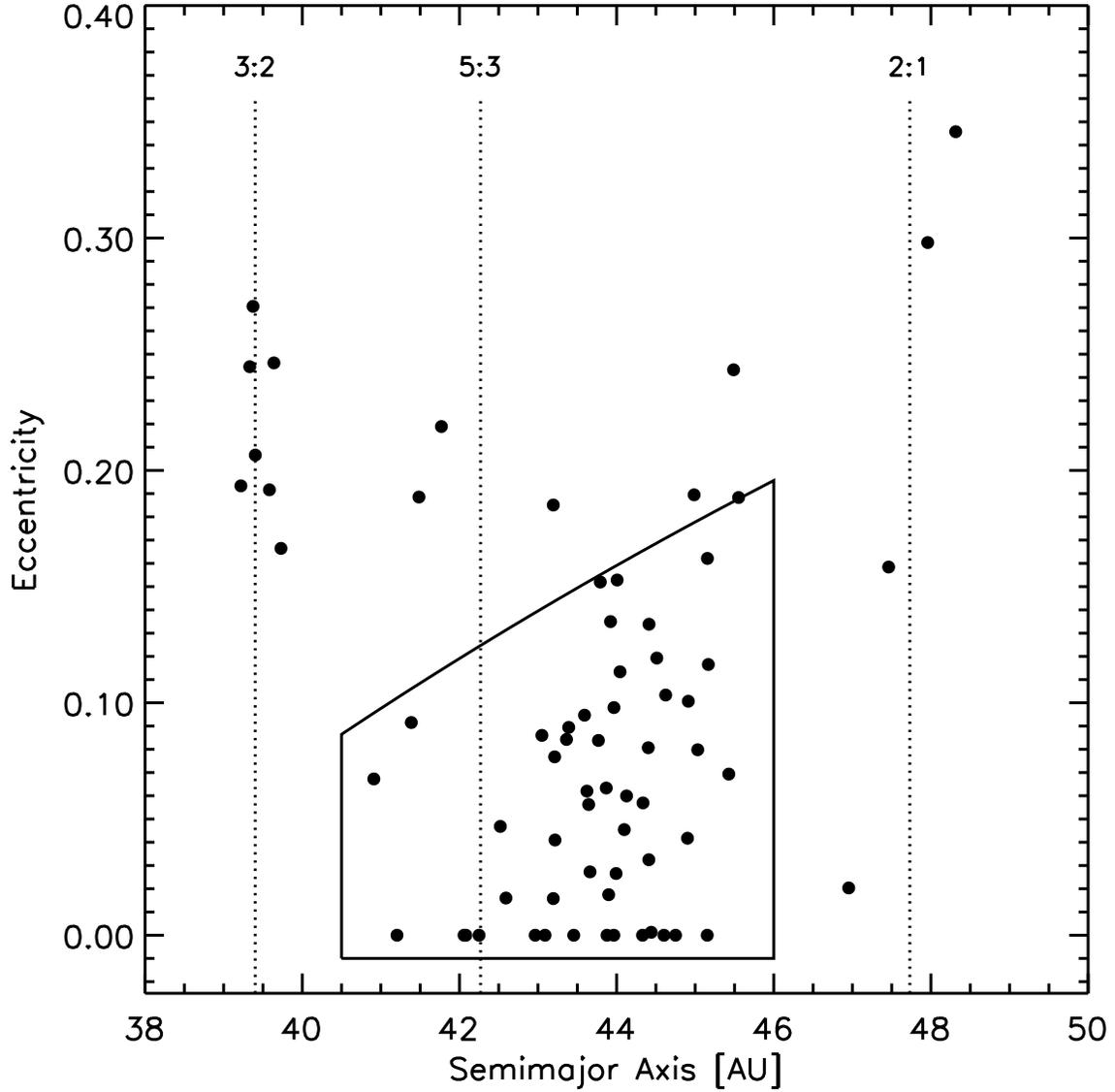}{6in}{0}{80}{80}{-250}{-100}
\figcaption{Eccentricity vs. semimajor axis of all KBOs discovered
in this work with semimajor axes $a < 50$ AU.  Note that few objects
were found in the 3:2 resonance compared to previous studies.  The
area enclosed by a solid line indicates our criteria for selecting
Classical KBOs, semimajor axes $40.5 \mbox{ AU} < a < 46 \mbox{ AU}$
and perihelia $q' > 37$ AU. \label{evsa1}}
\end{figure}

With only first opposition observations, the inclination and
heliocentric distance at discovery can be well determined for nearly
all KBOs, as depicted in Figure~\ref{ivsr-opp1}.  We find that the
semimajor axis and eccentricity determinations are less reliable but
are usually good enough to classify the objects as either Classical,
Resonance or Scattered KBOs, as depicted in Figure~\ref{evsa-opp1}.
We find that 6 out of 36 (17\%) of the objects exhibit orbital changes
large enough for their dynamical classification to change from the
first opposition to the second opposition.  Randomly rejecting 17\% of
our sample (to simulate misclassification) does not significantly
change the results.  In addition, rejection of all but the
multi-opposition objects does not significantly change our results; as
expected, the total number of KBOs estimated decreased by a factor
$\sim 2$ and error bars increased by a factor $\sim \sqrt{2}$ due to
the sample size reduction.  The eccentricity and semimajor axes of all
objects with $a<50$ AU (this includes all Classical KBOs) appear in
Figure~\ref{evsa1}.

In the next two sections, we use our observations to constrain three
fundamental quantities of the Classical KBOs: (1) the size
distribution index, $q$, (2) the half-width of the inclination
distribution, $i_{1/2}$, and (3) the total number of CKBOs larger than
100 km in diameter, $N_{\rm CKBOs}(D > 100 \mbox{ km})$.  The
quantities $i_{1/2}$ and $q$ are uncorrelated, as the observable
constraining $i_{1/2}$ is the inclination distribution and the
observable constraining $q$ is the absolute magnitude distribution.
However, $N_{\rm CKBOs}(D > 100 \mbox{ km})$ is a function of both $q$
and $i_{1/2}$ as a steeper size distribution or thicker inclination
distribution will each allow more bodies to be present.  In the
maximum likelihood simulations that follow, the ideal case would be to
constrain $q$, $i_{1/2}$ and $N_{\rm CKBOs}(D > 100 \mbox{ km})$ and
estimate errors in one simulation, however, this is difficult
computationally.  Therefore, we find the best-fit values of the three
parameters in a single simulation, but estimate the errors on the
parameters in two simulations, one that estimates the $q$-$N_{\rm
CKBOs}(D > 100 \mbox{ km})$ joint errors and one that estimates the
$i_{1/2}$-$N_{\rm CKBOs}(D > 100 \mbox{ km})$ joint errors.  We then
combine the two simulation results in quadrature to determine the
errors on $N_{\rm CKBOs}(D > 100 \mbox { km})$.

\section{The Size Distribution of the Classical KBOs}

We estimate the size distribution of the KBOs from our data in two
ways.  The first is a simple estimate made directly from the
distribution of ecliptic KBO apparent magnitudes (CLF).  The second is
a model which simulates the discovery characteristics of our survey
through the use of a maximum likelihood model constrained by the
absolute magnitude of the Classical KBOs.

\subsection{Cumulative Luminosity Function}
\label{clfsection}

We model the CLF with a power-law relation, $\log \Sigma = \alpha
(m_{R} - m_{0})$ (\S\ref{intro}).  The KBOs are assumed to follow a
differential power-law size distribution of the form $n(r) dr \propto
r^{-q} dr$, where $n(r) dr$ is the number of objects having radii
between $r$ and $r+dr$, and $q$ is the index of the size distribution.
Assuming albedo and heliocentric distance distributions that are
independent of KBO size, the simple transformation between the slope
of the CLF ($\alpha$) and the exponent of the size distribution ($q$)
is given by (Irwin et al. 1995),
\begin{equation}
\label{qeq}
q = 5 \alpha + 1.
\end{equation}
Under these assumptions, the size distribution can be estimated
directly from the CLF.

\begin{figure}
\plotfiddle{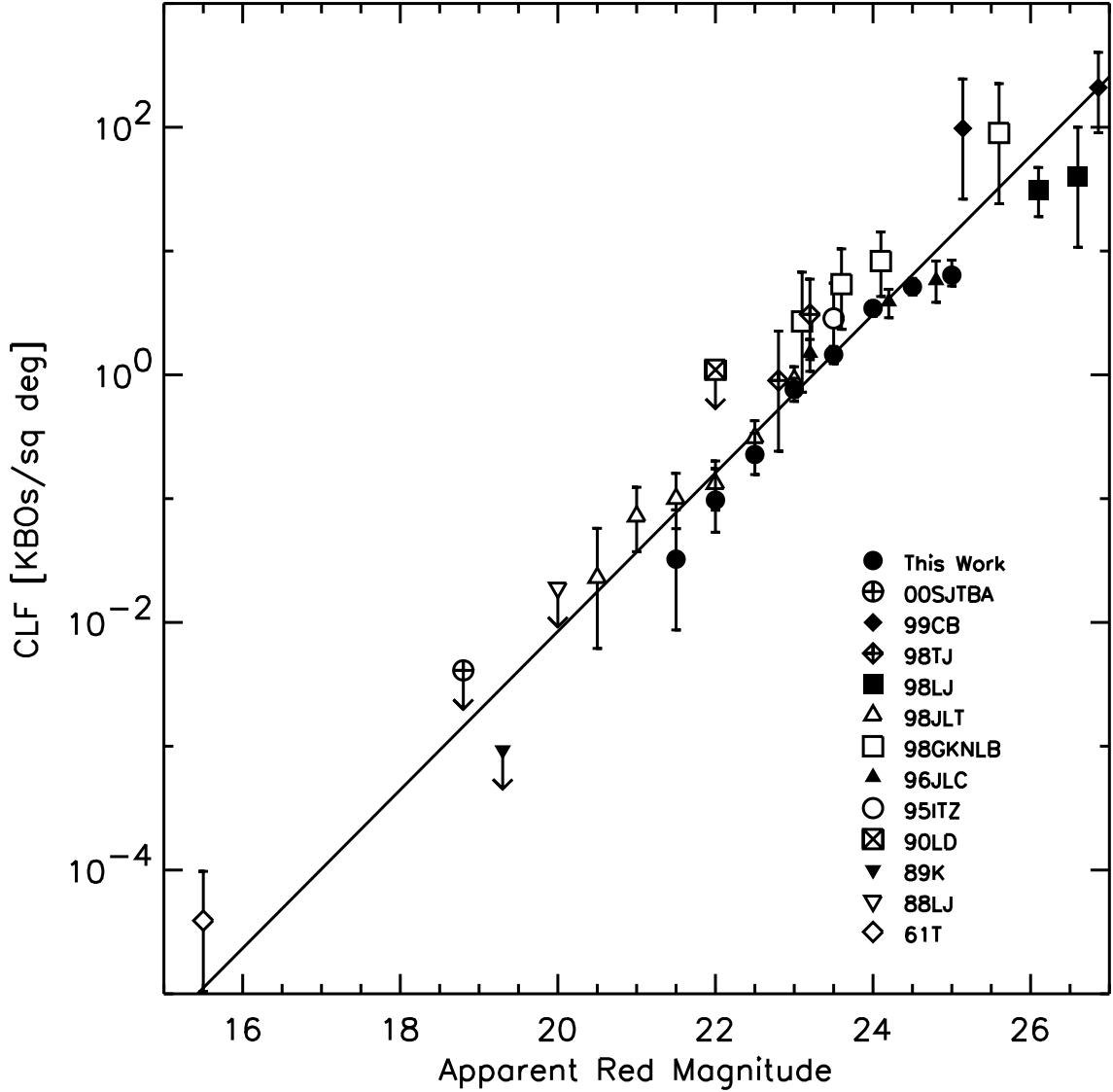}{6in}{0}{80}{80}{-250}{-100} \figcaption{Our
measurement of the Cumulative Luminosity Function (CLF), which
represents the number of KBOs $\mbox{deg}^{-2}$ near the ecliptic
(filled circles) brighter than a given apparent red magnitude.  Other
points are previous works (see text for abbreviations), with arrows
denoting upper limits.  The line represents a fit to our data alone,
yielding $\alpha = 0.63 \pm 0.06$, corresponding to $q = 4.15 \pm 0.3$
assuming the the albedo and heliocentric distance distributions are
independent of the size distribution.  Abbreviations are as follows:
00SJTBA is Sheppard et al. (2000), 99CB is Chiang \& Brown (1999),
98GKNLB is Gladman et al. (1998), 98JLT is Jewitt, Luu \& Trujillo
(1998), 98LJ is Luu \& Jewitt (1998), 98TJ is Trujillo \& Jewitt
(1998), 96JLC is Jewitt, Luu \& Chen (1996), 95ITZ is Irwin, Tremaine
\& \.{Z}ytkow (1995), 90LD is Levison \& Duncan (1990), 89K is Kowal
(1989), 88LJ is Luu \& Jewitt (1988), and 61T is Tombaugh (1961).
\label{clf}}
\end{figure}

We estimated the CLF by multiplying the detection statistics from the
observed distribution of object brightnesses by the inverse of the
detection efficiency.  We assumed Poisson detection statistics, with
error bars indicating the interval over which the integrated Poisson
probability distribution for the observed number of objects contains
68.27\% of the total probability (identical to the errors derived by
Kraft, Burrows \& Nousek 1991).  This is nearly equal to the Gaussian
case for all data points resulting from more than a few detections.
We have included all 74 KBOs discovered in our 37.2 sq deg of ecliptic
fields in the estimate of the CLF.  This includes the lost objects, as
the CLF is simply a count of the number of bodies discovered at a
given apparent magnitude.  Our results appear in Figure~\ref{clf},
with other published KBO surveys.  All observations were converted to
$R$-band if necessary assuming $V-R = 0.5$ for KBOs (Luu \& Jewitt
1996), and error bars were computed assuming Poisson detection
statistics. The data point of Cochran et al. (1995) near $m_{R} = 28$
was omitted because of major uncertainties about its reliability
(Brown, Kulkarni \& Liggett 1997, cf. Cochran et al.  1998).  Early
photographic plate surveys (Tombaugh 1961, portions of Luu \& Jewitt
1988, and Kowal 1989) have unproven reliability at detecting faint
slow-moving objects, and plate emulsion variations and defects make
accurate photometric calibration difficult.  The photographic plate
survey data were not used in our analysis.

The CLF points are highly correlated with one another, resulting in a
heavy weighting of the bright object data points.  Thus, we fitted the
Differential Luminosity Function (DLF) instead.  We plot the DLF data
points at the faint end of the bin, representing the modal value in
that bin.  Very small bin sizes were chosen (0.1 magnitudes) to negate
binning effects incurred from averaging the detection efficiency
(Equation~\ref{cfhtefffunc}) over a large magnitude range.  For any
non-zero CLF slope, $\alpha$, the DLF and CLF slopes are equal due to
the exponential nature of the CLF.  The DLF was modelled by evaluating
the Poisson probability of detecting the observed DLF given a range of
$m_{0}$ and $\alpha$, with the maximum probability corresponding to
our best-fit values.  Error bars were determined by finding the
contours of constant joint probability for $m_{0}$ and $\alpha$
enclosing 68.27\% of the total probability, a procedure similar to
that used below for the maximum likelihood simulation.  Computations
from this procedure are summarized in Table~\ref{clfdlfcomp}.  We find
that the slope of the CLF is $\alpha = 0.64^{+0.11}_{-0.10}$ with
$m_{0} = 23.23^{+0.15}_{-0.20}$, which corresponds to $q = 4.2 \pm
0.5$ from Equation~\ref{qeq}.  We also fitted the CLF by applying the
maximum likelihood method described by Gladman et al. (1998) to our
data, which yields statistically identical results to the binned DLF
procedure: $\alpha = 0.63 \pm 0.06$ and $m_{0} =
23.04^{+0.08}_{-0.09}$, corresponding to $q = 4.2 \pm 0.3$.  The
maximum-likelihood method provides slightly better signal-to-noise,
and is independent of binning effects.  However, it does not provide a
visualization of the data, as presented for the DLF fit.  We adopt the
maximum-likelihood procedure as our formal estimate of the CLF slope.
Both methods estimating the size distribution are in statistical
agreement with the more detailed analysis presented in the next
section.

The best-fit $\alpha = 0.63$ magnitude distribution was compared with
the observed magnitude distribution using a Kolmogorov-Smirnov test
(Press et al. 1992), producing a value of $D = 0.13$.  If the model
and the data distributions were identical, a deviation greater than
this would occur by chance 12\% of the time.  Thus, our linear model is
not a perfect fit, but it is statistically acceptable.

\subsection{Maximum Likelihood Simulation}

We now present more detailed analysis of the size distribution.  Since
we model the detection statistics of an assumed population, we choose
to model the 49 Classical KBOs discovered on the ecliptic as they are
numerically dominant in the observations and their orbital parameters
are more easily modelled than other KBO classes.  Our selection
criteria for CKBOs are perihelion $q' > 37 \mbox{ AU}$ and $40.5
\mbox{ AU} < a < 46 \mbox{ AU}$.  Given the size of an object and its
orbital parameters, we can compute its position, velocity, and
brightness, allowing a full ``Monte-Carlo'' style analysis of the bias
effects of our data collection procedures.  The apparent brightness
was computed from:
\begin{equation}
\label{mreq}
m = m_{\odot} - 2.5 \log(p \Phi(\alpha') r^{2}) + 2.5 \log(2.25 \times
10^{16} R^{2} \Delta^{2}),
\end{equation}
where $\alpha'$ is the phase angle of the object, $\Phi(\alpha')$ is
the Bowell et al. (1989) phase function, geometric albedo is given by
$p$, $r$ is the object radius in kilometers, $R$ is the heliocentric
distance, and $\Delta$ is the geocentric distance, both in AU (Jewitt
\& Luu 1995).  The apparent red magnitude of the Sun was taken to be
$m_{\odot} = -27.1$.  For this work, we assume $p = 0.04$, consistent
with a Centaur-like albedo (Jewitt \& Luu 2000).  We neglect phase
effects (setting $\Phi(\alpha') \equiv 1$) since the maximum phase
angle of an object at $R = 40$ AU within 1.5 hours of opposition is
$\alpha' = 0.55\arcdeg$.  This corresponds to $\Phi(\alpha') = 0.91$,
a change in brightness of only 0.09 magnitudes, which is less than
other uncertainties in the data.

This apparent brightness is used in a biasing-correction procedure
(Trujillo, Jewitt \& Luu 2000 and Trujillo 2000), summarized here:
\begin{enumerate}
\item A model distribution of KBOs is assumed (described in
Table~\ref{qmodel}).
\item KBOs are drawn randomly from the model distribution.
\item For each KBO, the apparent speed and ecliptic coordinates are
computed from the equations of Sykes \& Moynihan (1996, a sign error
was found in Equation 2 of their text and corrected), and compared to
the observed fields and speed criteria.
\item The apparent magnitude is computed from Equation~\ref{mreq}.
\item The efficiency function (Equation~\ref{cfhtefffunc}) and our
field area covered are used to determine if the simulated object would
be ``detected'' in our survey.
\item A histogram of the detection statistics for the simulated
objects is constructed, logarithmically binned by object size for the
size distribution model and binned by inclination for the
inclination-distribution model.  Binning effects were negligible due
to small bin choice.
\item Steps 1-6 are repeated until the number of detected simulated
objects is at least a factor 10 greater than the number of observed
objects in each histogram bin (typically requiring a sample of $10^{6}
< N < 10^{8}$ simulated objects, depending on the observed
distribution).
\item The likelihood of producing the observed population from the
model is estimated by assuming that Poisson detection statistics ($P =
\frac{\mu^{n}}{n!}  \exp\left({-\mu}\right)$) apply to each histogram
bin, where $\mu$ represents the expected number of simulated objects
``discovered'' given the number of objects simulated and $n$
represents the true number of KBOs observed.  Thus, the observed size
distribution, calculated from Equation~\ref{mreq}, is used to
constrain the $q$ model, and the observed inclination distribution is
used to constrain the $i$ model (\S~\ref{idist}).
\end{enumerate}
These steps are repeated for each set of model parameters in order to
estimate the likelihood of producing the observations for a variety of
models.

\begin{figure}
\plotfiddle{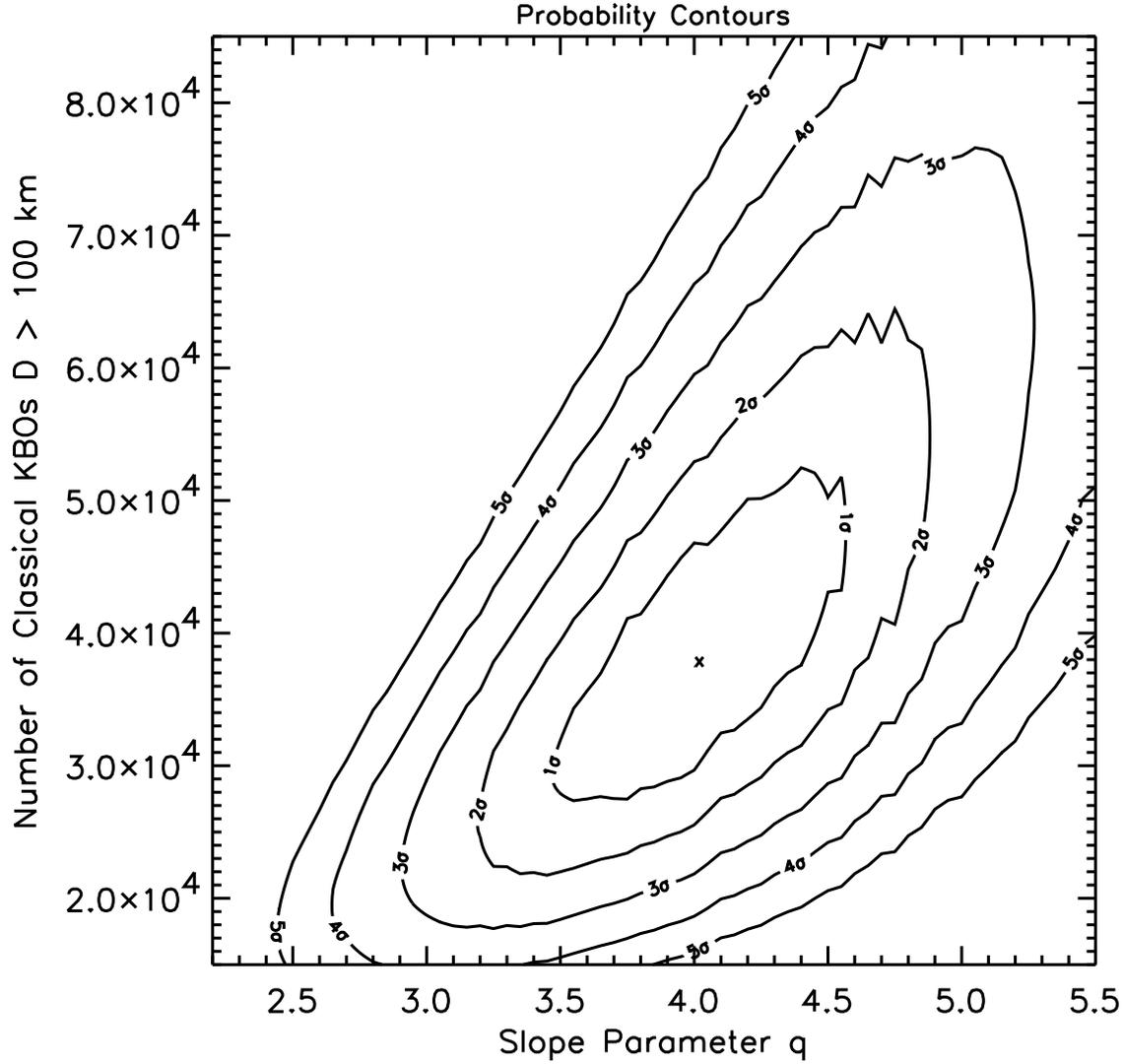}{6in}{0}{80}{80}{-250}{-100} \figcaption{The
maximum likelihood simulation of the size distribution power-law
exponent.  Contours of constant likelihood ($1\sigma$, $2\sigma$,
... $5\sigma$) are shown for a model with differential size
distribution $q$ (x-axis) and total number of objects greater than 100
km in diameter $N(D > 100 \mbox{ km})$ (y-axis).  The maximum
likelihood parameters (denoted by an x) occur at $q = 4.0$ and $N_{\rm
CKBOs}(D > 100 \mbox{ km}) = 3.8 \times 10^{4}$.
\label{kboqdist}}
\end{figure}

\begin{figure}
\plotfiddle{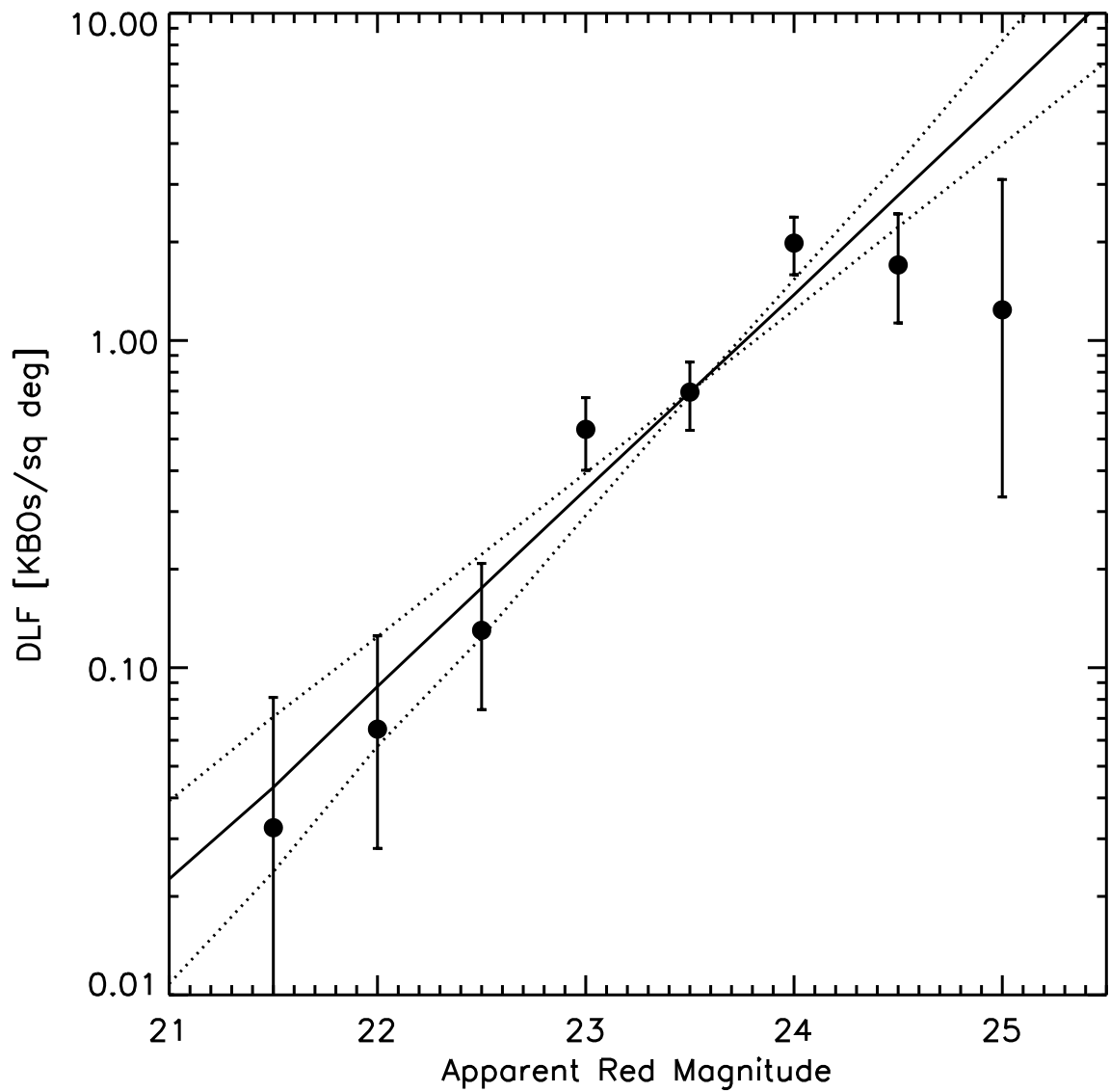}{6in}{0}{80}{80}{-250}{-100} \figcaption{The
Differential Luminosity Function (DLF), equal to the number of KBOs
$\mbox{deg}^{-2}$ near the ecliptic (filled circles).  Three different
models of the observed magnitude distribution are plotted from our
maximum likelihood model (lines), representing the expected DLF for
the $+1\sigma$ (dotted), best-fit (solid), and $-1\sigma$ (dotted)
cases of $q = $ 3.5, 4.0, and 4.6, respectively.
\label{dlfq}}
\end{figure}

For the size distribution analysis, we take our best-fit model of the
width of the inclination distribution (Half-Width $i_{1/2} =
20^{\circ}$, as estimated in the next section), and vary the size
distribution index $q$, and the total number of objects $N_{\rm
CKBOs}(D > 100 \mbox{ km})$.  Model parameters are summarized in
Table~\ref{qmodel} and results appear in Figure~\ref{kboqdist}.  Our
best-fit values are
\begin{displaymath}
\begin{array}{rcll}
q        & = & 4.0^{+0.6}_{-0.5}  & (1 \sigma) \mbox{ and} \\
         & = & 4.0^{+1.3}_{-2.1} & (3 \sigma), \\
         & \mbox{and} & & \\
N_{\rm CKBOs}(D > 100 \mbox{ km}) & = & 3.8^{+2.0}_{-1.5} \times 10^{4} & (1 \sigma) \mbox{ and} \\
                                  & = & 3.8^{+5.4}_{-2.7} \times 10^{4} & (3 \sigma), \\
\end{array}
\end{displaymath}
where the errors for $N_{\rm CKBOs}(D > 100 \mbox{ km})$ have been
combined in quadrature from the results of the $q$ and $i_{1/2}$ fits,
as described at the end of \S~\ref{recovery}.  The values for $q$ are
consistent with previously published works (Table~\ref{sizetable}) and
the $q$ derived from the CLF data in the simple model
(Equation~\ref{qeq}).  The results are consistent with the
distribution of large ($D > 150$ km) main-belt asteroids ($q = 4.0$,
Cellino, Zappal\'{a}, \& Farinella 1991) and rock crushed by
hypervelocity impacts ($q=3.4$, Dohnanyi 1969).  In addition, the
scenario where the cross-sectional area (and thus optical scattered
light and thermal emission) is concentrated in the largest objects
($q<3$, Dohnanyi 1969) is ruled out at the $>2 \sigma$ ($>95.4\%$
confidence) level.  Our results are also consistent with Kenyon \& Luu
(1999) who simulate the growth and velocity evolution of the Kuiper
Belt during the formation era in the Solar System.  They find several
plausible models for the resulting size distribution, all of which
have $q \approx 4$.  In Figure~\ref{dlfq} we plot the best-fit model
CKBO distribution with the observed DLF to demonstrate the expected
results from different size distributions.

The magnitude distribution expected from the maximum likelihood model
was compared to the observed magnitude distribution, as was done for
the CLF-derived magnitude distribution in \S~\ref{clfsection}.  The
Kolmogorov-Smirnov test produced $D = 0.17$; a greater deviation would
occur by chance 11\% of the time.

In our Classical KBO maximum likelihood simulation, we have ignored
possible contributions of the 7 lost KBOs, since their orbital classes
are not known.  However, including them in the simulations by assuming
circular orbits at the heliocentric distance of discovery results in
statistically identical resulted for $q$, and the expected 7/49 rise
in $N_{\rm CKBOs}(D > 100 \mbox{ km})$.

\section{Inclination Distribution of the Classical KBOs}
\label{idist}

\begin{figure}
\plotfiddle{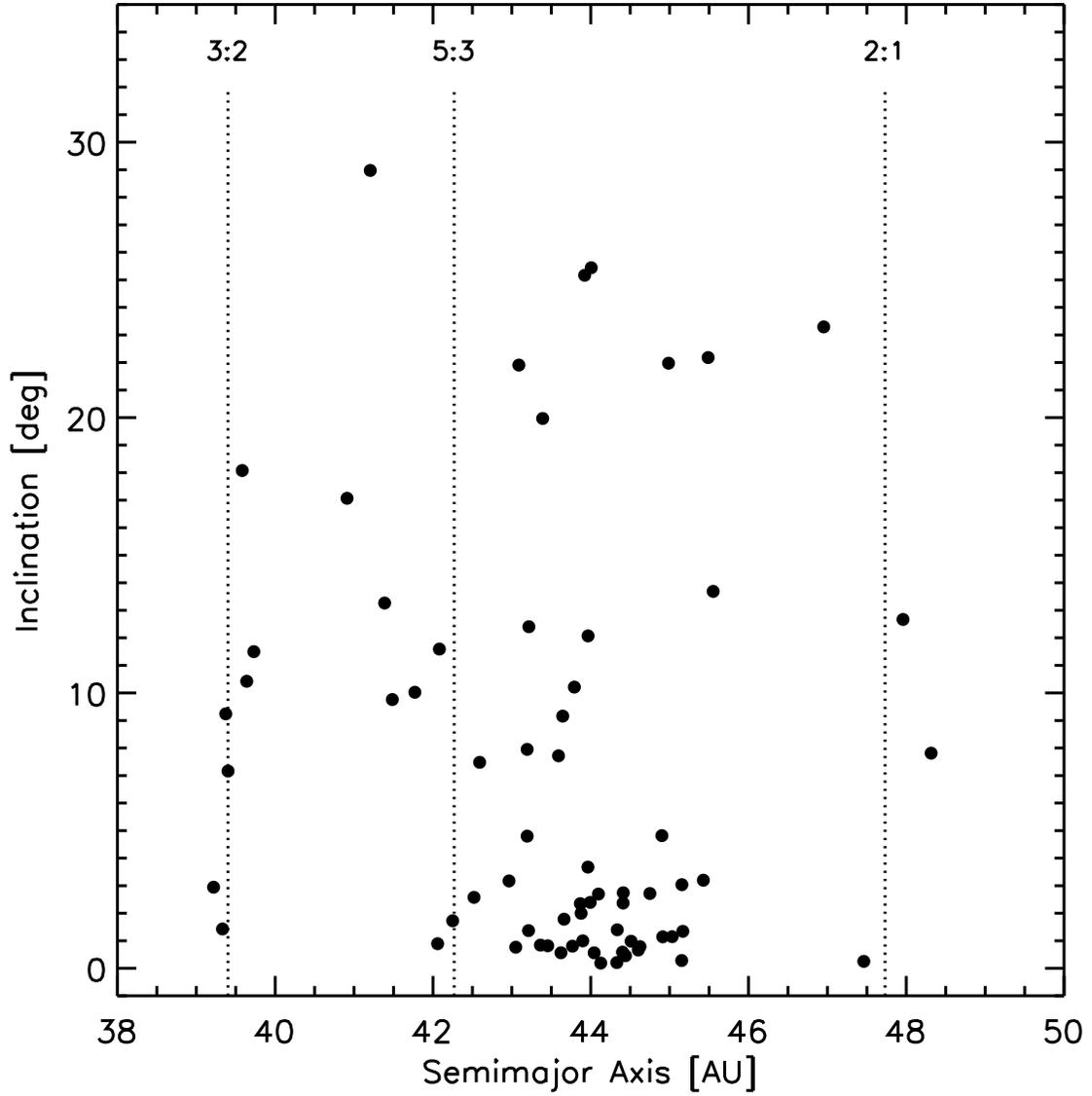}{6in}{0}{80}{80}{-250}{-100}
\figcaption{Inclination vs. semimajor axis of all KBOs discovered
in this work with semimajor axes $a < 50$ AU.
\label{ivsa1}}
\end{figure}

\begin{figure}
\plotfiddle{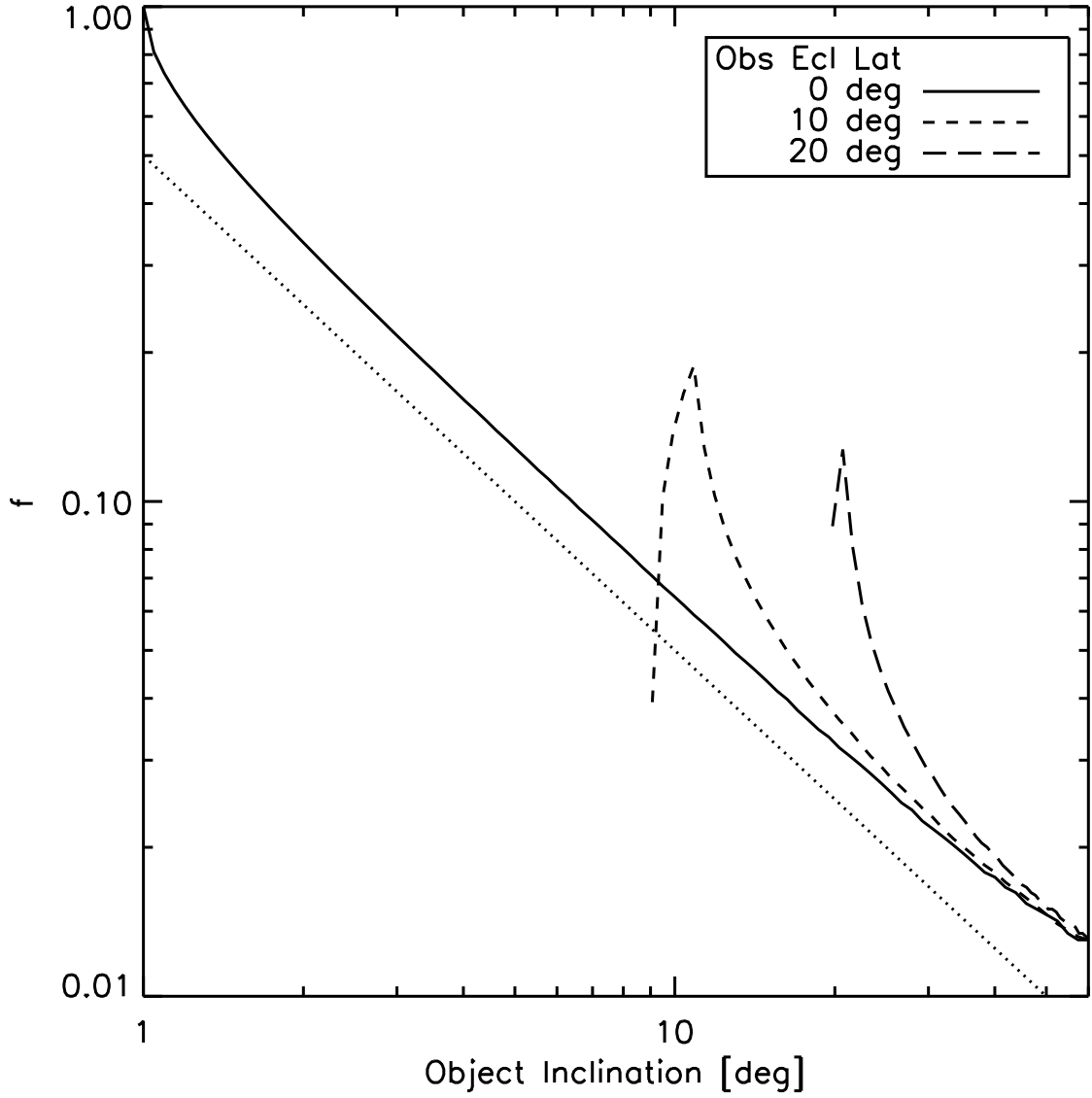}{6in}{0}{80}{80}{-250}{-100}
\figcaption{The fraction $f$ of an orbit spent within $\pm1^{\circ}$
(solid line), $10^{\circ} \pm 1^{\circ}$ (short dashed line), and
$20^{\circ} \pm 1^{\circ}$ (long dashed line) of the ecliptic, as a
function of object inclination $i$.  The dotted line has a slope of
1/$i$.
\label{ifig}}
\end{figure}

The dynamical excitation of the Kuiper Belt is directly related to the
inclination distribution of the KBOs.  We present the inclinations of
the CKBOs found in the CFHT survey in Figure~\ref{ivsa1}.
Assuming heliocentric observations, a KBO in circular
orbit follows
\begin{equation}
\label{beta}
\sin \beta = \sin i \sin f
\end{equation}
where $\beta$ is the heliocentric ecliptic latitude, $0 < i <
90^{\circ}$ is the inclination, and $0 < f < 360^{\circ}$ represents
the true anomaly of the object's orbit with $f = 0$ and $180\arcdeg$
representing the ecliptic plane crossing (the longitude of perihelion
is defined as 0 in this case).  Using Equation~\ref{beta}, we plot the
fraction of each orbit spent at various ecliptic latitudes as a
function of $i$ (Figure~\ref{ifig}).  This plot demonstrates two
trends concerning the ecliptic latitude of observations $\beta_{\rm
obs}$.  First, high-inclination objects are a factor 3--4 times more
likely to be discovered when $\beta_{\rm obs} \sim i$ than when
observing at low ecliptic latitudes ($\beta_{\rm obs} < i$).  Second,
the number of expected high-inclination objects drops precipitously,
roughly as $1/i$, once $i > 1.5 \beta_{\rm obs}$ (Jewitt, Luu \& Chen
1996).

These facts led us to observe at three different ecliptic latitudes
($0^{\circ}$, $10^{\circ}$ and $20^{\circ}$) to better sample the
high-inclination objects.  During two observation periods (Sep 1999
and Mar 2000) care was made to interleave the ecliptic fields with the
off-ecliptic fields on timescales of $\sim 30$ minutes.  This
technique provides immunity to drift in the limiting magnitude which
might otherwise occur in response to typical slow changes in the
seeing through the night.  The results for the robust, interleaved
fields matched those for the seeing-corrected Feb 1999 fields where
fields were interleaved on much longer timescales of $\sim 3$ hours.
Accordingly, we combined the data sets from all epochs to improve
signal-to-noise.  In the next sections, we analyse the inclination
distribution using two techniques to demonstrate the robustness of our
method.

\subsection{Simple Inclination Model}
\label{simplemodel}

First, since fields were imaged at three different ecliptic latitudes,
the surface density of objects at each latitude band
($\Sigma(0\arcdeg)$, $\Sigma(10\arcdeg)$ and $\Sigma(20\arcdeg)$) can
directly yield the underlying inclination distribution.  In our simple
model, we generate an ensemble of inclined, circular orbits drawn from
a Gaussian distribution centered on the ecliptic, and having a
characteristic Half-Width of $i_{1/2}$.  The probability of drawing a
KBO with inclination between $i$ and $i+di$ is given by
\begin{equation}
\label{ieq}
P(i)di = \frac{1}{\sigma \sqrt{2 \pi}} \exp \left(\frac{-i^{2}}{2 \sigma^{2}} \right) di,
\end{equation}
where $\sigma = i_{1/2} / \sqrt{2 \ln 2}$.  Using this relation, and
Equation~\ref{beta}, we simulate the expected values of
$\Sigma(0\arcdeg)$, $\Sigma(10\arcdeg)$ and $\Sigma(20\arcdeg)$ for
various $i_{1/2}$.  These are compared to two ratios measured from our
observations, $R(10^{\circ},0^{\circ}) \equiv \Sigma(10\arcdeg) /
\Sigma(0\arcdeg)$ and $R(20^{\circ},0^{\circ}) \equiv
\Sigma(20\arcdeg) / \Sigma(0\arcdeg)$.  Results appear in
Table~\ref{simplei}, and demonstrate that the characteristic
half-width of the inclination distribution in the Kuiper Belt is
$i_{1/2} \sim 17\arcdeg^{+10\arcdeg}_{-4\arcdeg}$ ($1 \sigma$ =
68.27\% confidence).  This simple model does not use the observed
inclination distribution of the individual objects, merely the surface
density of objects found at each ecliptic latitude, thus we have
combined all objects from all KBO classes into this estimate.

\subsection{Full Maximum Likelihood Inclination Model}
\label{fullmodel}

\begin{figure}
\plotfiddle{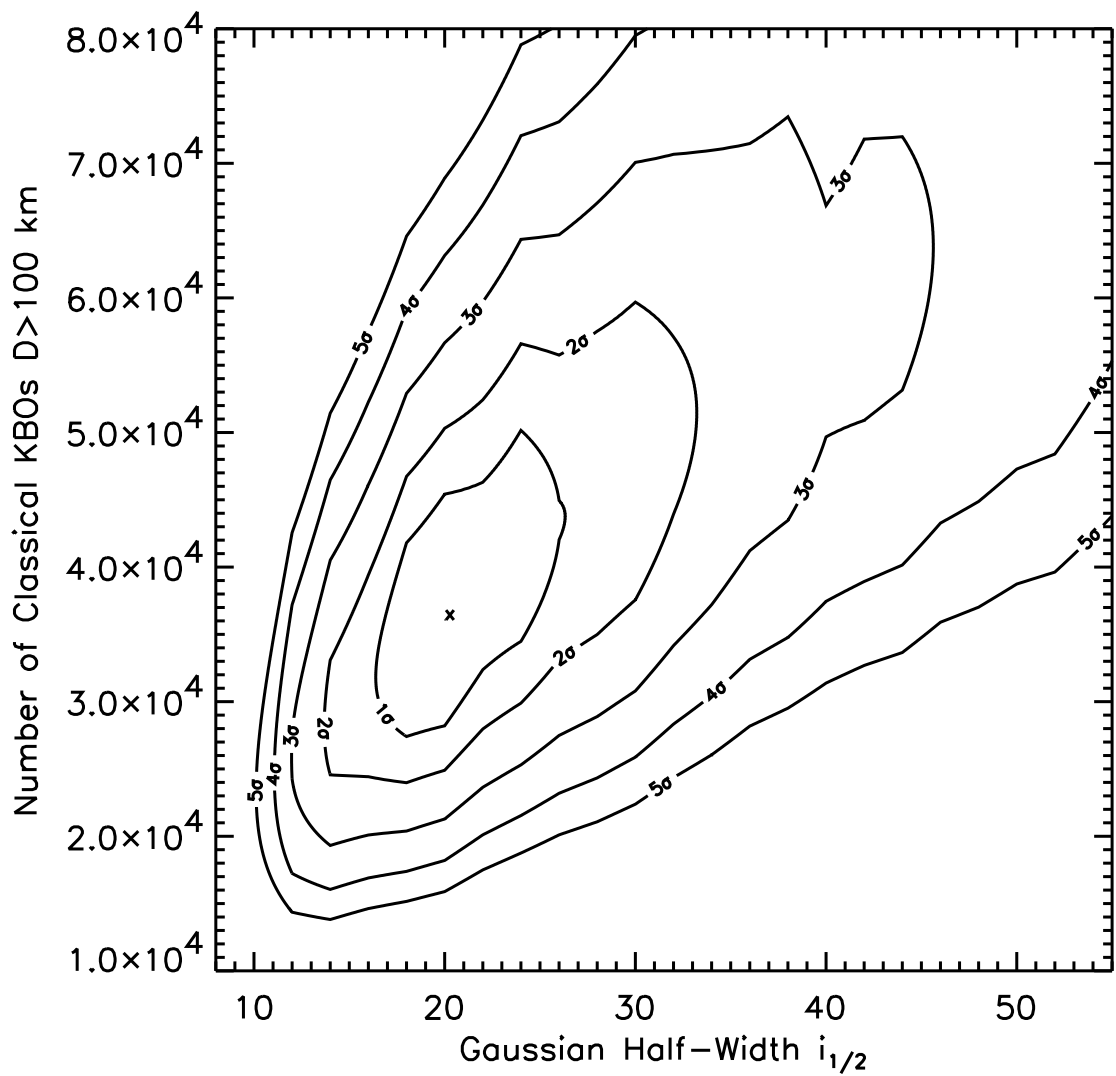}{6in}{0}{80}{80}{-250}{-100} \figcaption{The
maximum likelihood simulation.  Contours of constant likelihood
($1\sigma$, $2\sigma$, ... $5\sigma$) are shown for a model with
Gaussian half-width $i_{1/2}$ (x-axis) and total number of CKBOs with
diameters greater than 100 km $N_{\rm CKBOs}(D > 100 \mbox { km})$
(y-axis).  The maximum likelihood occurs at $N_{\rm CKBOs}(D > 100
\mbox{ km}) = 3.8 \times 10^{4}$ and $i_{1/2} = 20\arcdeg$.
\label{iprobs}}
\end{figure}

\begin{figure}
\plotfiddle{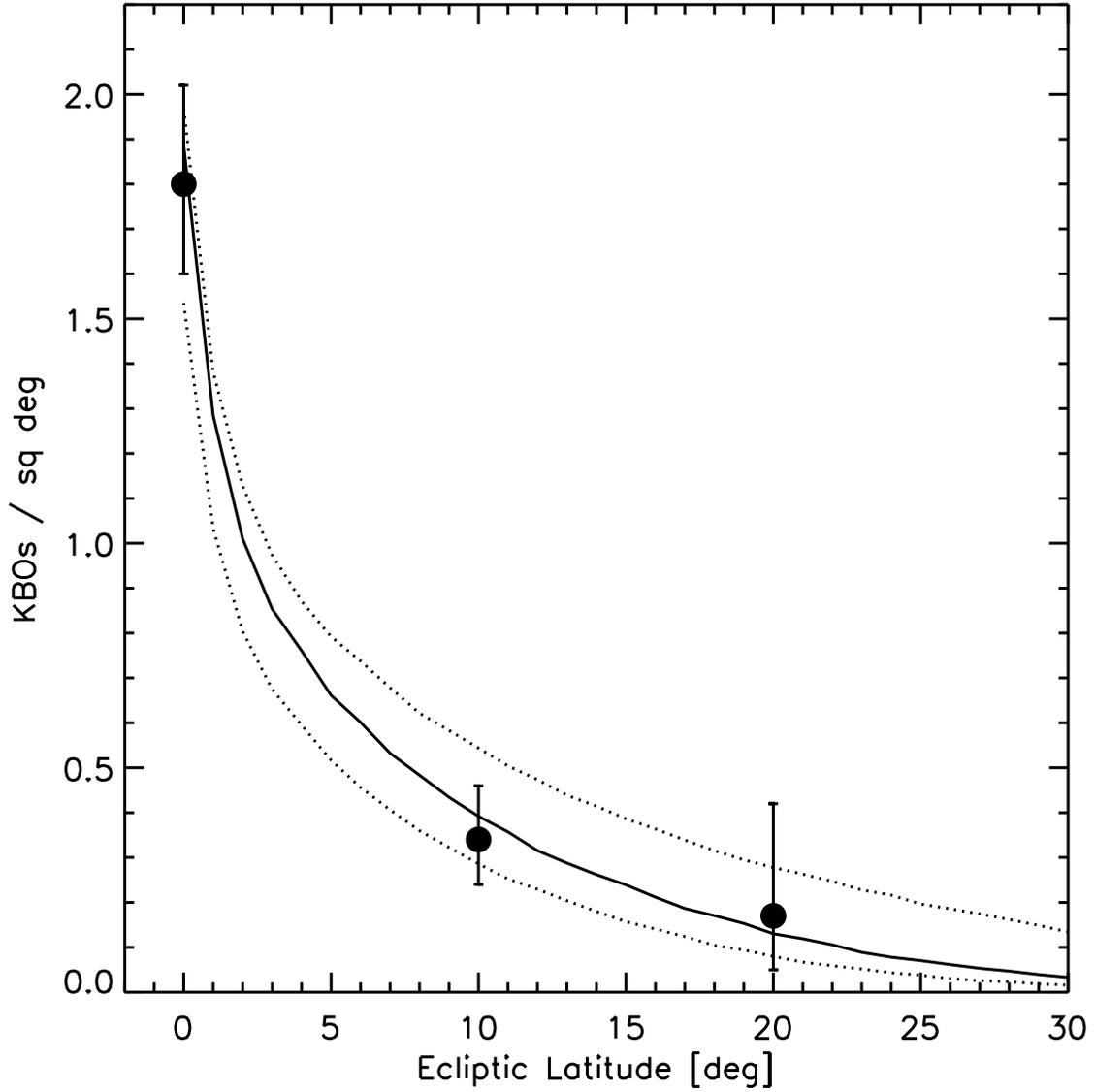}{6in}{0}{80}{80}{-250}{-100}
\figcaption{Surface density of KBOs brighter than $m_{R} = 23.7$
vs. ecliptic latitude.  The solid line represents the best fit
$i_{1/2} = 20\arcdeg^{+6\arcdeg}_{-4\arcdeg}$ CKBO model while the
dotted lines represent the $1\sigma$ errors.  The CKBO model has been
multiplied by the observed KBO/CKBO ratio ($86/49 = 1.76$) for display
purposes, to simulate the surface density of the more numerous KBOs.
\label{isigplot}}
\end{figure}

Second, we use the maximum likelihood model described in
\S\ref{clfsection}.  We list the parameters of the model in
Table~\ref{fulli}.  This model encompasses the additional constraint
of the observed inclination distribution, as well as the parallactic
motion of the Earth and KBO orbital motion to produce more realistic
results.  Results appear in Figure~\ref{iprobs}, with $N_{\rm CKBOs}(D
> 100 \mbox { km})$ representing the number of CKBOs with diameters
greater than 100 km.  The maximum likelihood occurs at
\begin{displaymath}
\begin{array}{rcll}
i_{1/2}  & = & 20\arcdeg^{+6\arcdeg}_{-4\arcdeg}  & (1 \sigma) \mbox{ and} \\
         & = & 20\arcdeg^{+26\arcdeg}_{-8\arcdeg} & (3 \sigma), \\
         & \mbox{and} & & \\
N_{\rm CKBOs}(D > 100 \mbox{ km}) & = & 3.8^{+2.0}_{-1.5} \times 10^{4} & (1 \sigma) \mbox{ and} \\
                                  & = & 3.8^{+5.4}_{-2.7} \times 10^{4} & (3 \sigma), \\
\end{array}
\end{displaymath}
where the errors for $N_{\rm CKBOs}(D > 100 \mbox{ km})$ have been
estimated from the $i_{1/2}$ and $q$ fits, combined in quadrature, as
described at the end of \S~\ref{recovery}.  This maximum likelihood
model is consistent with the simple model described in
\S\ref{simplemodel}.  In Figure~\ref{isigplot}, we plot the observed
surface density of objects as a function of ecliptic latitude and
compare these data to our best-fit models.  This illustrates the
fundamental fact that even though the true inclination distribution of
the KBOs is very thick ($i_{1/2} \approx 20\arcdeg$), the surface
density drops off quickly with ecliptic latitude, reaching half the
ecliptic value at an ecliptic latitude of $\beta \approx 3\arcdeg$
($\Sigma(3\arcdeg) / \Sigma(0\arcdeg) < 0.5$).

The functional form of the inclination distribution cannot be well
constrained by our data.  However, the best-fit Gaussian distribution
was compared to a flat-top (``top-hat'') inclination distribution,
with a uniform number of objects in the $0\arcdeg < i < 30\arcdeg$
range.  The Gaussian and flat-top models were equally likely to
produce the observed distribution in the 65\% confidence limit ($< 1
\sigma$).  A Gaussian model multiplied by $\sin(i)$ was also tried but
could be rejected at the $>3 \sigma$ level because it produced too few
low-inclination objects.  We also tested the best-fit model presented
by Brown (2001), consisting of two Gaussians multiplied by $\sin(i)$,
\begin{equation}
[a \exp(\frac{-i^{2}}{2 \sigma_{1}^2}) + (1-a) \exp(\frac{-i^{2}}{2
\sigma_{2}^{2}})] \sin i,
\end{equation}
where $a = 0.93$, $\sigma_{1} = 2.2\arcdeg$, and $\sigma_{2} =
18\arcdeg$, and found it equally compatible with our single Gaussian
model (Equation ~\ref{ieq}).  Because the Gaussian model was the
simplest model that fit the observed data well, we chose it to derive
the following velocity dispersion results.

We first find the mean velocity vector of all the simulated best-fit
CKBOs, $\vec{\bar{v}}$, in cylindrical coordinates (normal vectors
$\hat{r}$, $\hat{\theta}$, and $\hat{z}$ representing the radial,
longitudinal and vertical components respectively).  The mean velocity
vector $\vec{\bar{v}}$ is consistent with a simple Keplerian rotation
model at $R \approx 46$ AU.  We then compute the relative velocity of
each KBO from this via $\vert \vec{\bar{v}} - \vec{v_{i}} \vert$,
where $\vec{v_{i}}$ is the velocity dispersion contribution of the
$i$th KBO.  We find the resulting root-mean-square (RMS) velocity
dispersion of the $\hat{r}$, $\hat{\theta}$, and $\hat{z}$ components
to be equal to $\Delta v_{r} = 0.51$ km/s, $\Delta v_{\theta} = 0.50$
km/s, and $\Delta v_{z} = 0.91$ km/s, combining in quadrature for a
total velocity dispersion of $\Delta v = \sqrt{\Delta v_{r}^{2} +
\Delta v_{\theta}^{2} + \Delta v_{z}^{2}} = 1.16$ km/s.  An error
estimate of the velocity dispersion can be found by following a
similar procedure for the $i_{1/2}$ = 16\arcdeg and 26\arcdeg ($\pm 1
\sigma$) models, yielding $\Delta v = 1.16^{+0.25}_{-0.16}$ km/s.

\subsection{Inferred Mass}

The Kuiper Belt mass inferred from these results can be directly
calculated from the size distribution and the number of bodies
present.  For the best-fit $q=4.0$ size distribution, the mass of
CKBOs in bodies with diameters $D_{\rm min} < D < D_{\rm max}$ is
\begin{equation}
\label{masseq}
M(D_{\rm min},D_{\rm max}) = \frac{4}{3} \pi \rho \Gamma \ln(D_{\rm max}/D_{\rm min}),
\end{equation}
where $\rho$ is the bulk density of the object.  The normalization
constant $\Gamma$ is calculated from the results of our simulation,
\begin{equation}
\label{gammaeq}
\Gamma \approx 3.0 \times 10^{12} \mbox{ m}^{3} p_{R}^{-1.5} N(D > 100 \mbox{ km}),
\end{equation}
where $N(D > 100 \mbox{ km}) = 3.8 \times 10^{4}$ (\S\ref{fullmodel}),
yielding $\Gamma = 1.4 \times 10^{19} \mbox{ m}^{3}$ assuming $p_{R}
\equiv 0.04$.  The mass for $100 \mbox{ km} < D < 2000 \mbox{ km}$
then becomes
\begin{equation}
M(100 \mbox{ km}, 2000 \mbox{ km}) \approx 0.03 M_{\earth} \left(\frac{\rho}{1000 
\mbox{ kg} \mbox{ m}^{-3}}\right) \left(\frac{0.04}{p_{R}}\right)^{1.5},
\end{equation}
where $M_{\earth} = 6.0 \times 10^{24}$ kg is the mass of the earth.
The uncertainties on this value are considerable as the characteristic
albedo and density of the CKBOs are unknown.

\subsection{Comparison of the Classical KBOs to Other Dynamical Classes}

We found that the total number of CKBOs is given by $N_{\rm CKBOs}(D >
100 \mbox{ km}) = 3.8^{+2.0}_{-1.5} \times 10^{4}$.  This can be
compared to the other main dynamical populations (the Resonant and
Scattered KBOs) from our data.  Observational biases favor the
detection of the Plutinos over the Classical KBOs due to their closer
perihelion distance.  We found only 7 Plutinos (4 ecliptic and 3
off-ecliptic) so we can make only crude (factor $\sim 2$ statements)
about the true size of the population.  Thus, we use the results of
Jewitt, Luu \& Trujillo (1998) who estimate that the apparent fraction
of Plutinos ($P_{a}$) in the Kuiper Belt is enhanced relative to the
intrinsic fraction ($P_{i}$) by a factor $P_{a}/P_{i} \approx 2$ for
$q=4.0$ and $r_{\rm max} = 1000$ km.  Applying this correction to our
ecliptic observations (4 Plutinos and 49 Classical KBOs) indicates
that the total number of Plutinos larger than 100 km in diameter is
quite small,
\begin{equation}
N_{\rm Plutinos}(D > 100 \mbox{ km}) \approx \frac{4}{4+49}
\frac{P_{i}}{P_{a}} N_{\rm CKBOs} \approx 1400.
\end{equation}

The populations of the Plutinos and the 2:1 Resonant objects are
important measures of the resonance sweeping hypothesis (Malhotra
1995), which predicts equal numbers of objects in each resonance.
Since the 2:1 objects are systematically farther from the sun than the
Plutinos, the true Plutino/2:1 ratio is higher than the observed
ratio.  Jewitt, Luu \& Trujillo (2000) estimate the observed/true bias
correction factor to be $\approx 0.310$ for a survey similar to ours
($q = 4$ and $m_{R50} = 24.0$).  Only 2 of our objects (both found on
the ecliptic) are $< 0.5$ AU from the 2:1 Resonance, so we find the
Plutino/2:1 fraction is given by $(4/2) 0.310 \approx 0.6$.  Due to
the small number of bodies involved, this is only an order of
magnitude estimate.  Within the uncertainties, our observations are
consistent with the hypothesis that the 3:2 and 2:1 resonances are
equally populated.

The observational biases against the Scattered KBOs are considerable.
Trujillo, Jewitt \& Luu (2000) estimate the total population of the
Scattered KBOs to be $N_{\rm SKBOs}(D > 100 \mbox{ km}) =
3.1^{+1.9}_{-1.3} \times 10^{4}$, approximately equal to the
population of Classical KBOs derived from our data.  We summarize the
relative populations by presenting their number ratios:
\begin{equation}
\mbox{\it Classical} : \mbox{\it Scattered} : \mbox{\it Plutino} : \mbox{\it Resonant 2:1} = 1.0:0.8:0.04:0.07.
\end{equation}

\section{The Edge of the Classical Kuiper Belt}

\begin{figure}
\plotfiddle{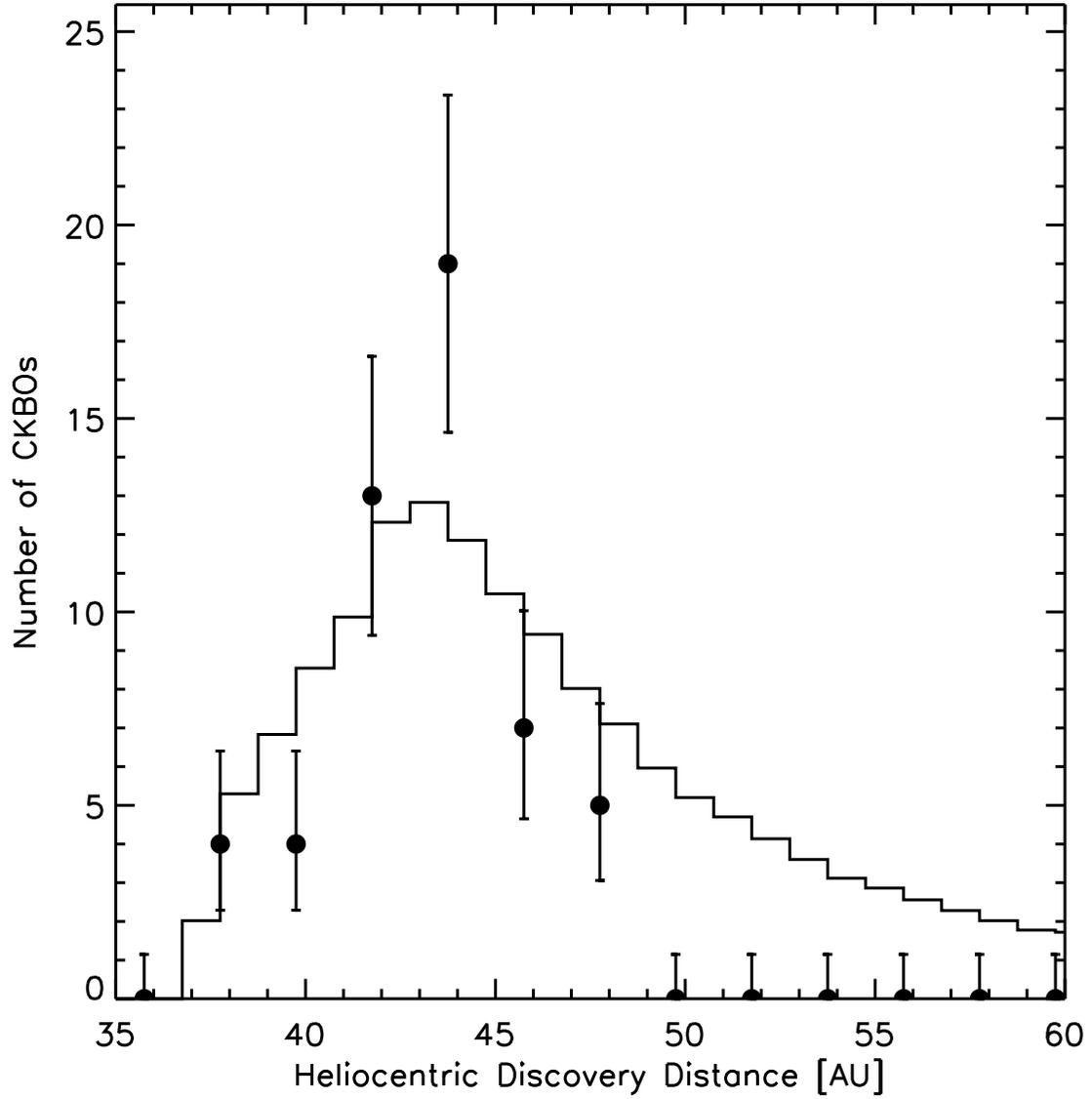}{6in}{0}{80}{80}{-250}{-100}
\figcaption{Observed heliocentric discovery distance (data points) and
expected discoveries assuming the best-fit untruncated CKBO model
(solid line).  Note the very sharp drop in discovery statistics
beginning at $\sim 46$ AU, violating the model.  This is consistent
with an outer edge to the Classical Kuiper Belt at 50 AU ($3
\sigma$). \label{edge}}
\end{figure}

We found no objects beyond heliocentric distance $R_{\rm obs} = 48.9$
AU.  There are two possibilities to explain this observation: (1) this
is an observational bias effect and the bodies beyond $R_{\rm obs}$
cannot be detected in our survey, or (2) there is a real change in the
physical or dynamical properties of the KBOs beyond $R_{\rm obs}$.  In
order to test these two possibilities, we compare the expected
discovery distance of an untruncated Classical Kuiper Belt to the
observations, as depicted in Figure~\ref{edge}.  This untruncated CKBO
distribution is identical to our best-fit model from
\S~\ref{fullmodel}, except $40.5 \mbox{ AU} < a < 200 \mbox{ AU}$,
instead of $40.5 \mbox{ AU} < a < 46 \mbox{ AU}$.  The total number of
bodies produced was considered a free parameter in this model.
Inspecting Figure~\ref{edge}, the absence of detections beyond 50 AU
is inconsistent with an untruncated model with $R^{-2}$ radial power
to the ecliptic plane surface density.  Assuming Poisson statistics
apply to our null detection beyond $R_{\rm max}$, the 99.73\% ($3
\sigma$) upper limit to the number of bodies ($\mu$) expected beyond
$R_{\rm max}$ can be calculated from $1-0.9973 = \exp(-\mu)$, yielding
$\mu = 5.9$ KBOs.  We found 49 ecliptic Classical KBOs inside the
$R_{\rm max}$ limit, so the $3 \sigma$ upper limit to the number
density of KBOs beyond $R_{\rm max}$ is $49/5.9 \approx 8$ times less
than the number density of Classical KBOs.  Although we have
constrained the outer edge by the heliocentric distance at discovery
$R$, which is a directly observable quantity, a dynamical edge would
be set by the semimajor axes ($a$) of the object orbits.  This
difference has little effect on our findings as the known CKBOs occupy
nearly circular orbits with median eccentricity $e = 0.08$ (the
calculated median is conservative as it includes only bodies with $e >
0$ to protect against short-arc orbits, which typically assume $e =
0$).  Since an untruncated distribution (1) is incompatible with our
data, we must conclude that scenario (2) applies --- there must be a
physical or dynamical change in the KBOs beyond $R_{\rm max}$.

There are several possible physical and dynamical mechanisms that
could produce the observed truncation of the belt beyond $R_{\rm max}
= 50$ AU (Jewitt, Luu \& Trujillo 1998): (1) the size distribution of
the belt might become much steeper beyond $R_{\rm max}$, putting most
of the mass of the belt in the smallest, undetectable objects; (2) the
size distribution could be the same ($q = 4$), but there might be a
dearth of large (i.e.  bright) objects beyond $R_{\rm max}$,
suggesting prematurely arrested growth; (3) the objects beyond $R_{\rm
max}$ may be much darker and therefore remain undetected; (4) the
eccentricity distribution could be lower in the outer belt, resulting
in the detection of fewer bodies; (5) the ecliptic plane surface
density variation with radial distance may be steeper than our assumed
$p = 2$; and (6) there is a real drop in the number density of objects
beyond $R_{\rm max}$.  We consider each of these cases in turn, and
their possible causes.

Detailed simulations of the growth of planetesimals in the outer Solar
System have not estimated the radial dependence of the formation
timescale (e.g.  Kenyon \& Luu 1999).  However, it is expected that
growth timescales should increase rapidly with heliocentric distance,
perhaps as $t \propto R^{3}$ (Wetherill 1989).  One could then expect
a reduction in the number of large objects beyond 50 AU, as per (2)
above, and a correspondingly steeper size distribution, as in (1), at
larger heliocentric distances.  However, with $t \propto R^{3}$, the
timescales for growth at $R = 41$ AU (inner edge) and $R = 50$ AU
(outer edge) are only in the ratio 1.8:1.  In addition, we observe no
correlation between size and semimajor axis among the Classical KBOs.

To test scenario (1), we took our untruncated best-fit model and
varied the size distribution index $q_{\rm out}$ for bodies with
semimajor axes $a > R_{\rm max}$, keeping the KBO mass across the
$R_{\rm max}$ boundary constant.  We then found the minimum $q_{\rm
out}$ consistent with our null detection beyond $R_{\rm max}$.  This
mass-conservation model is very sensitive to the chosen minimum body
radius $r_{\rm min}$, because for any $q_{\rm out} > 4$, most of the
mass is in the smallest bodies (Dohnanyi 1969).  The minimum
size-distribution index required as a function of $r_{\rm min}$
appears in Table~\ref{qedgetable}.  If mass is conserved for the
observable range of bodies, $r_{\rm min} = 50$ km, the observed edge
cannot be explained by a change in the size distribution unless $q>10$
($3 \sigma$), an unphysically large value.  For the conservative case
of $r_{\rm min} = 6$ km (roughly the size of cometary nuclei, Jewitt
1997), the observed edge could only be explained by $q > 5.6$ ($3
\sigma$) beyond $R_{\rm max}$.  We know of no population of bodies
with a comparably steep size distribution.  Thus, we conclude that the
observed edge is unlikely to be solely caused by a change in the size
distribution beyond $R_{\rm max}$.

A similar procedure was followed for possibility (2).  Here again, we
took our best-fit truncated model and extended it to large
heliocentric distances.  Then, $r_{\rm max}$ was varied to find the
largest value that could explain our null detection beyond $R_{\rm
max}$, keeping the total number density of objects with radii $r <
r_{\rm max}$ constant.  We found that $r_{\rm max} < 75$ km
($3\sigma$) was required beyond $R_{\rm max}$ to explain the observed
edge.  This is a factor $\sim 5$ smaller radius and a factor $\sim
150$ less volume than our largest object found within $R_{\rm max}$
(1999 $\rm CD_{158}$, $\sim 400$ km in radius).  Such a severe change
in the maximum object size beyond $R_{\rm obs}$ would have to occur
despite the fact that growth timescales vary by less than a factor of
$\sim 2$ over the observed Classical KBO range, as explained above.

One might also expect (3) to be true, as KBO surfaces could darken
over time with occasional resurfacing by collisions (Luu \& Jewitt
1996), and long growth timescales indicate long collision timescales
as well.  However, the geometric albedo would have to be $p< 0.008$, a
factor 5 lower than that of the CKBOs in our model, assuming a
constant number density of objects across the transition region.  We
are not aware of natural planetary materials with such low albedos.

The dynamical cases (4), a drop in the eccentricity distribution, and
(5), a steeper ecliptic plane density index, can also be rejected.
Even an extreme change in the eccentricity distribution cannot explain
our observations.  Lowering eccentricity from $e = 0.15$ (a high value
for the Classical KBOs) to $e = 0$ results in a perihelion change from
42.5 AU to 50 AU for an object with semimajor axis 50 AU.  Such a
change corresponds to a 0.7 magnitude change in perihelion brightness,
and to a factor 2.8 change in the surface density of objects expected
from our $\alpha = 0.63$ CLF.  This model is rejected by our
observations at the $> 5 \sigma$ level.  The variation in ecliptic
plane surface density with respect to heliocentric distance was
assumed to follow a power law with index $p = 2$ in our model.
However, even a large increase to $p = 5$ would result in a reduction
in surface density of a factor 2.7 in the 41 AU to 50 AU range, which
can also be rejected as the cause of our observed edge at the $> 5
\sigma$ level.

Since scenarios (1) through (5) seem implausible at best, we conclude
that the most probable explanation for the lack of objects discovered
beyond $R_{\rm max}$ is (6), the existence of a real, physical
decrease in object number density.  There have been few works
considering mechanisms for such truncation.  The 2:1 mean-motion
Neptune resonance at $a \sim 47.8$ AU is quite close to the observed
outer edge of the belt. However, given the Neptune resonance sweeping
model (Malhotra 1995), the resonance could not cause an edge.  The
sweeping theory predicts that the 2:1 resonance should have passed
through the Classical Kuiper Belt as Neptune's orbit migrated outwards
to its present semimajor axis.  Thus, the KBOs interior to the current
2:1 resonance ($a \approx 47.8$ AU) could have been affected by this
process, but an edge at $R_{\rm max}$ cannot be explained by such a
model.  Ida, Larwood \& Burkert (2000) simulate the effect of a close
stellar encounter on the Kuiper Belt, suggesting that KBO orbits
beyond 0.25--0.3 times the stellar perihelion distance would be
disrupted and ejected for a variety of encounter inclinations.  Thus,
an encounter with a solar mass star with perihelion at $\sim 200$ AU
might explain the observed edge.  Such encounters are implausible in
the present solar environment but might have been more common if the
sun formed with other stars in a dense cluster.

\section{Constraints on a Distant Primordial Kuiper Belt}

While our observations indicate a dearth of objects beyond 50 AU, it
is also possible that a ``wall'' of enhanced number density exists at
some large $R \gtrsim 100$ AU distance, as suggested by Stern (1995).
We know that the Kuiper Belt has lost much mass since formation
because the present mass is too small to allow the observed objects to
grow in the age of the solar system.  Kenyon \& Luu (1999) found that
the primordial Kuiper Belt mass in the $30 \mbox{ AU} < R < 50 \mbox{
AU}$ region could have been some $\sim 10 M_{\oplus}$, compared to the
$\sim 0.1 M_{\oplus}$ we see today.  Stern (1995) also speculated that
the primordial surface density may be present at large heliocentric
distances.  We model this primordial belt as analogous to the CKBOs in
terms of eccentricity, inclination and size distribution, but
containing a factor of 100 more objects and mass per unit volume of
space.  These objects would be readily distinguishable from the rest
of the objects in our sample as they would have low eccentricities
characteristic of the CKBOs ($e < 0.25$) yet would have very large
semimajor axes ($a > 90$ AU).  Since we have discovered no such
``primordial'' objects, Poisson statistics state that the $3 \sigma$
upper limit to the sky-area number density of primordial KBOs is 5.9
in 37.2 sq deg, or 0.16 primordial KBOs $\mbox{deg}^{-2}$.  We
constrain the primordial KBOs by allowing the inner edge of the
population, $a_{\rm min}$, to vary outwards, while keeping the outer
edge fixed at 250 AU.  We find that $a_{\rm min} = 130$ AU coincides
with the $3 \sigma$ limit on the inner edge of the belt, nearly at the
extreme distance limit of our survey.  An object discovered at our
survey magnitude limit $m_{R50} = 23.7$ at this distance would have
diameter $D \approx 1800$ km (approximately 25\% smaller than Pluto)
assuming a 4\% albedo.

\section{Summary}

New measurements of the Kuiper Belt using the world's largest CCD
mosaic array provide the following results in the context of our
Classical KBO model.

(1) The slope of the differential size distribution, assumed to be a
    power law, is $q = 4.0^{+0.6}_{-0.5}$ ($1 \sigma$).  This is
    consistent with accretion models of the Kuiper Belt (Kenyon \& Luu
    1999).  This distribution implies that the surface area, the
    corresponding optical reflected light and thermal emission are
    dominated by the smallest bodies.

(2) The Classical KBOs inhabit a thick disk with Half-Width
    $20\arcdeg^{+6\arcdeg}_{-4\arcdeg}$ ($1 \sigma$).

(3) The Classical KBOs have a velocity dispersion of
    $1.16^{+0.25}_{-0.16}$ km/s.

(4) The population of Classical KBOs larger than 100 km in diameter
    $N_{\rm CKBOs}(D > 100 \mbox{ km}) = 3.8^{+2.0}_{-1.5} \times
    10^{4}$ ($1 \sigma$).  The corresponding total mass of bodies with
    diameters between 100 km and 2000 km is $M(100 \mbox{ km}, 2000
    \mbox{ km}) \sim 0.03 M_{\earth}$ , assuming geometric red albedo
    $p_{R} \equiv 0.04$ and bulk density $\rho \equiv 1000 \mbox{ kg}
    \mbox { m}^{-3}$.

(5) The approximate population ratios of the Classical, Scattered, 3:2
    Resonant (Plutinos) and 2:1 Resonant KBOs are 1.0:0.8:0.04:0.07.

(6) The Classical Kuiper Belt has an outer edge at $R = 50$ AU.  This
    edge is unlikely to be due to a change in the physical properties
    of the CKBOs (albedo, maximum object size, or size distribution).
    The edge is more likely a real, physical depletion in the number
    of bodies beyond $\sim 50$ AU.

(7) There is no evidence of a primordial (factor 100 density increase)
    Kuiper Belt out to heliocentric distance $R = 130$ AU.

\acknowledgements

We thank Dave Tholen and Brian Marsden for providing orbits and
ephemerides.  We appreciate the vital observational assistance
provided by Scott Sheppard and his help with astrometric measurements.
We thank David Woodworth, Ken Barton, Lisa Wells and Christian Viellet
for help at the CFH telescope.  We are grateful for the assistance of
John Dvorak, Chris Merrick, Lance Amano, Paul DeGrood and Farren
Herron-Thorpe at the University of Hawaii 2.2 m telescope.  A NASA
grant to DCJ provided financial support for this project.


\newpage

\newpage

\newpage

\newpage

\newpage

\newpage

\newpage

\newpage

\newpage

\newpage

\newpage

\newpage

\newpage

\clearpage

\begin{center}
\begin{deluxetable}{ccc}
\small
\tablewidth{0pt}
\tablecaption{CFHT Survey Parameters}
\tablehead{\colhead{Quantity} & \colhead{CFHT 3.6m}}
\startdata
Focal Ratio & f/4 \nl
Instrument & CFHT 12k x 8k Mosaic \nl
Plate Scale [arc sec/pixel] & 0.206 \nl
North-South extent [deg] & 0.47 \nl
East-West extent [deg] & 0.70 \nl
Field Area [$\rm deg^{2}$] & 0.330 \nl
Total Area [$\rm deg^{2}$] & 73 \nl
Integration Time [sec] & 180 \nl
Readout Time [sec] & 60 \nl
$m_{R50}$\tablenotemark{a} & 23.7 \nl
$\theta$\tablenotemark{b} [arc sec] & 0.7--1.1 \nl
Filter & $R$ \nl
Quantum Efficiency & 0.75 \nl
\enddata
\tablenotetext{a}{the red magnitude at which detection efficiency reaches half of the maximum efficiency}
\tablenotetext{b}{the typical Full Width at Half Maximum of stellar sources
for the surveys.}
\label{obs}
\end{deluxetable}
\end{center}

\clearpage

\begin{center}
\begin{deluxetable}{lcc}
\small
\tablewidth{0pt}
\tablecaption{CFHT 12k Photometric Calibration}
\tablehead{\colhead{chip} & \colhead{$m_{0} \pm \sigma$} & $N$}
\startdata
Feb 1999 \nl
00 & $25.63 \pm 0.03$ & 2 \nl
01 & $25.74 \pm 0.08$ & 4 \nl
02 & $25.78 \pm 0.08$ & 7 \nl
03 & $26.07 \pm 0.10$ & 8 \nl
04 & $26.08 \pm 0.08$ & 6 \nl
05 & $26.06 \pm 0.09$ & 4 \nl
06 & not used        &   \nl
07 & $25.79 \pm 0.04$ & 4 \nl
08 & $26.00 \pm 0.10$ & 17\nl
09 & $25.98 \pm 0.05$ & 6 \nl
10 & $26.03 \pm 0.01$ & 2 \nl
11 & $26.13 \pm 0.04$ & 2 \nl
Sep 1999 and Mar 2000 \nl
00 (04 in Feb 1999) & $26.08 \pm 0.05$ & 6 \nl
01 & $25.74 \pm 0.04$ & 4 \nl
02 & $25.78 \pm 0.05$ & 7 \nl
03 \tablenotemark{a} & $25.99 \pm 0.09$ & 3 \nl
04 (05 in Feb 1999) & $26.06 \pm 0.08$ & 4 \nl
05 (11 in Feb 1999) & $26.13 \pm 0.04$ & 2 \nl
06 \tablenotemark{a} & $25.56 \pm 0.02$ & 5 \nl
07 & $25.79 \pm 0.03$ & 4 \nl
08 & $26.00 \pm 0.10$ & 17\nl
09 & $25.98 \pm 0.04$ & 6 \nl
10 \tablenotemark{a} & $25.49 \pm 0.06$ & 2 \nl
11 (03 in Feb 1999) & $26.07 \pm 0.08$ & 8 \nl
\enddata
\tablecomments{Zero points were consistent between observing runs,
however, 3 chips were replaced and several of the remaining chips were
shifted in position and renumbered after the Feb 1999 run.}
\tablenotetext{a}{New chip added after Feb 1999 run.}
\label{cfhtphot}
\end{deluxetable}
\end{center}

\clearpage

\begin{center}
\begin{deluxetable}{ccccc}
\small
\tablewidth{0pt}
\tablecaption{CFHT Survey Efficiency}
\tablehead{\colhead{Quantity} & \colhead{Good} & \colhead{Medium} & \colhead{Poor} & \colhead{Global}}
\startdata
Median PSF FWHM [\arcsec]       & 0.76       & 0.90       & 1.07       & 0.84       \nl
PSF FWHM Range  [\arcsec]       & 0.56--0.80 & 0.80--0.99 & 1.00--1.40 & 0.56--1.40 \nl
$e_{\rm max}$                   & 0.83       & 0.83       & 0.83       & 0.83       \nl
$m_{R50}$                       & 24.01      & 23.64      & 23.35      & 23.74      \nl
$\sigma$                        & 0.29       & 0.38       & 0.47       & 0.48       \nl
Fields Imaged                   & 95         & 89         & 49         & 233        \nl
\enddata
\label{effs}
\end{deluxetable}
\end{center}

\clearpage

\begin{center}
\begin{deluxetable}{cccl}
\small
\tablewidth{0pt}
\tablecaption{Classical KBO Size Distribution Model Parameters}
\tablehead{\colhead{Symbol} & \colhead{Value} & \colhead{Distribution} & \colhead{Description}}
\startdata
$a$                         & 40.5 -- 46 AU     & $n(a) da \sim a^{1-p} da$ & semimajor axis\nl
$p$\tablenotemark{a}        & 2               & --- & semimajor axis power\nl
$e$                         & 0 -- 0.25        & uniform & eccentricity \nl
$q'$                        & $q' > 37 \mbox{ AU}$ & --- & perihelion distance\nl 
$i$                         & 0 -- 90 deg     & Gaussian, $i_{1/2}$ half-width & inclination distribution\nl
$i_{1/2}$                   & 20 deg          & --- & Half-Width of the $i$ distribution\nl
$\omega$                    & 0 -- 360 deg    & uniform & argument of perihelion\nl
$\Omega$                    & 0 -- 360 deg    & uniform & longitude of the ascending node\nl
$M$                         & 0 -- 360 deg    & uniform & mean anomaly \nl
$r$                         & 50 -- 1000 km   & $n(r) dr \sim r^{-q} dr$ & radius\nl
$q$                         & fitted          & --- & size distribution index\nl
$p_{R}$                     & 0.04            & --- & geometric red albedo \nl
$N_{\rm CKBOs}(D > 100 \mbox{ km})$                         & fitted & --- & number of CKBOs with $D > 100$ km \nl
---                         & 20              & --- & number of radius bins (log intervals) \nl
---                         & 50 --- 1000 km  & --- & radius bin range \nl
\enddata
\tablenotetext{a}{In the circular orbit case, $p$ corresponds to the
power of the decrease in ecliptic plane surface density $\Sigma_{\rm
ecl}$ as a function of heliocentric distance $R$, $\Sigma_{\rm ecl} \sim R^{-p}$.}
\label{qmodel}
\end{deluxetable}
\end{center}

\clearpage

\begin{center}
\begin{deluxetable}{cccl}
\small
\tablewidth{0pt}
\tablecaption{Selected Size Distribution Measurements of the KBOs}
\tablehead{
\colhead{$m_{R}$ Range} & \colhead{Number}      & \colhead{$q$} & \colhead{Reference} \\
\colhead{of Discovery}   & \colhead{of KBOs found} & \colhead{}     & \colhead{} }
\startdata
21.1 -- 24.6 & 86 & $ 4.0^{+0.6}_{-0.5} $ & This Work \nl
25.5 -- 27.2\tablenotemark{a} &  2 & $ 3.6 \pm 0.1 $ & Chiang \& Brown 1999\tablenotemark{b} \nl
23.8 -- 26.7 &  6 & $ 3.7 \pm 0.2 $ & Luu \& Jewitt 1998\tablenotemark{b} \nl
23.0 -- 25.8 &  5 & $ 4.8^{+0.5}_{-0.6} $ & Gladman et al. 1998\tablenotemark{b} \nl
20.6 -- 23.0 & 13 & $ 4.0 \pm 0.5 $ & Jewitt, Luu \& Trujillo 1998\tablenotemark{b} \nl
\enddata
\tablenotetext{a}{V magnitude}
\tablenotetext{b}{Extrapolated from CLF slope, $\alpha$, via Equation~\ref{qeq}.}
\label{sizetable}
\end{deluxetable}
\end{center}

\clearpage

\begin{center}
\begin{deluxetable}{cccc}
\small
\tablewidth{0pt}
\tablehead{\colhead{Ecliptic Latitude [deg]} & \colhead{0} & \colhead{10\tablenotemark{a}} & \colhead{20}}
\tablecaption{Simple Inclination Model}
\startdata
Number of Fields                             & 119                    & 94                     & 19                    \nl
Field Area [sq deg]                          & 37.2                   & 29.8                   & 6.0                   \nl
Number of KBOs                               & 74                     & 11                     & 1                     \nl
Surface Density [\#/sq deg]\tablenotemark{b} & $1.80^{+0.22}_{-0.20}$ & $0.34^{+0.12}_{-0.10}$ & $0.17^{+0.25}_{-0.12}$\nl
$R(10^{\circ},0^{\circ})$ and $R(20^{\circ},0^{\circ})$ & ---         & $0.19^{+0.07}_{-0.06}$ & $0.09^{+0.13}_{-0.07}$\nl
$i_{1/2}$ [deg]                              &                        & $14^{+6}_{-3}$         & $19^{+20}_{-7}$        \nl
\enddata
\tablenotetext{a}{Results for ecliptic latitude $\beta = +10\arcdeg$ are
consistent with those of $\beta = -10\arcdeg$, so were combined.}
\tablenotetext{b}{Error bars were computed assuming Poisson detection
statistics (Kraft, Burrows \& Nousek 1991).}
\label{simplei}
\end{deluxetable}
\end{center}

\clearpage

\begin{center}
\begin{deluxetable}{cccl}
\small
\tablewidth{0pt}
\tablecaption{KBO Inclination Model Parameters}
\tablehead{\colhead{Symbol} & \colhead{Value} & \colhead{Distribution} & \colhead{Description}}
\startdata
$a$        & 40.5 -- 46 AU & $n(a)da \sim a^{1-p} da$ & semimajor axis\nl
$p$\tablenotemark{a}  & 2 & --- & semimajor axis power \nl
$e$        & 0 -- 0.25     & uniform & eccentricity \nl
$q'$                        & $q' > 37 \mbox{ AU}$ & --- & perihelion distance\nl 
$i$        & 0 -- 90 deg   & Gaussian, $i_{1/2}$ half-width & inclination distribution\nl
$i_{1/2}$  & fitted        & --- & Half-Width of the $i$ distribution\nl
$\omega$   & 0 -- 360 deg  & uniform & argument of perihelion\nl
$\Omega$   & 0 -- 360 deg  & uniform & longitude of the ascending node\nl
$M$        & 0 -- 360 deg  & uniform & mean anomaly \nl
$r$        & 50 -- 1000 km & $n(r) dr \sim r^{-q} dr$& radius\nl
$q$        & 3.7           & --- & slope parameter \nl
$p_{R}$    & 0.04          & --- & red albedo \nl
$N_{\rm KBOs}(D > 100 \mbox{ km})$        & fitted        & --- & number of CKBOs with $D > 100$ km \nl
---        & 45            & --- & number of inclination bins \nl
---        & 0\arcdeg --- 90\arcdeg & --- & inclination bin range \nl 
\enddata
\tablenotetext{a}{In the circular orbit case, $p$ corresponds to the
power of the decrease in ecliptic plane surface density $\Sigma_{\rm
ecl}$ as a function of heliocentric distance $R$, $\Sigma_{\rm ecl} \sim R^{-p}$.}
\label{fulli}
\end{deluxetable}
\end{center}

\clearpage

\begin{center}
\begin{deluxetable}{ll}
\small
\tablewidth{2.5in}
\tablecaption{Minimum $q_{\rm out}$ Needed to Explain Observed Edge}
\tablehead{\colhead{$r_{\rm min}$\tablenotemark{a}} & \colhead{$q_{\rm out}$}}
\startdata
50.0 & 10 \nl
25.0 & 7.4 \nl
12.5 & 6.2 \nl
 6.3 & 5.6 \nl
\enddata
\tablenotetext{a}{Minimum radius [km] for which mass is conserved
across the edge boundary.}
\label{qedgetable}
\end{deluxetable}
\end{center}

\end{document}